\documentclass[letterpaper,twocolumn,10pt]{article}
\usepackage{usenix2019_v3}

\usepackage{epsfig,endnotes}
\usepackage{subfig}
\usepackage{epstopdf}
\usepackage{arydshln}
\usepackage{booktabs}
\usepackage{amsfonts} 
\usepackage{graphicx} 
\usepackage{xcolor} 
\usepackage{enumitem}
\usepackage{eso-pic}
\usepackage{wrapfig} 
\usepackage{algorithmicx}
\usepackage[ruled]{algorithm}
\usepackage{algpseudocode} 
\usepackage{amsmath}
\usepackage{xspace}
\usepackage{epstopdf}
\usepackage{listings}
\usepackage{color}
\usepackage{tikz}
\usepackage{amsmath}
\usepackage{xargs}                      
\usepackage{xcolor}

\newcommand{\F}{\mathbb{F}}
\newcommand{\Z}{\mathbb{Z}}
\newcommand{\fTPM}{{\sf fTPM}}

\makeatletter
\newcommand\footnoteref[1]{\protected@xdef\@thefnmark{\ref{#1}}\@footnotemark}
\makeatother

\usepackage{hyperref}
    \definecolor{linkcolor}{rgb}{0,0,0.25}
    \definecolor{citecolor}{rgb}{0,0.4,0}
    \definecolor{urlcolor}{rgb}{0,0,0.65}
    \hypersetup{pdfpagemode=UseNone,pdfstartview=FitH,colorlinks=true, linkcolor=linkcolor, urlcolor=urlcolor, citecolor=citecolor}

\definecolor{dkgreen}{rgb}{0,0.6,0}
\definecolor{gray}{rgb}{0.5,0.5,0.5}
\definecolor{mauve}{rgb}{0.58,0,0.82}
\usepackage{todo}

\usepackage{authblk}
\author[1]{Daniel Moghimi}
\author[1]{Berk Sunar}
\author[1, 2]{Thomas Eisenbarth}
\author[3]{Nadia Heninger}
\affil[1]{Worcester Polytechnic Institute, Worcester, MA, USA}
\affil[2]{University of L\"ubeck, L\"ubeck, Germany}
\affil[3]{University of California, San Diego, CA, USA}

\begin{document}

\title{\textsc{TPM-Fail}: TPM meets Timing and Lattice Attacks}

\maketitle

\begin{abstract}

Trusted Platform Module (TPM) serves as a hardware-based root of trust that protects cryptographic keys from privileged system and physical adversaries. 
In this work, we perform a black-box timing analysis of TPM 2.0 devices deployed on commodity computers.
Our analysis reveals that some of these devices feature secret-dependent execution times during signature generation based on elliptic curves.
In particular, we discovered timing leakage on an Intel firmware-based TPM as well as a hardware TPM.
We show how this information allows an attacker to apply lattice techniques to recover 256-bit private keys for ECDSA and ECSchnorr signatures.
On Intel fTPM, our key recovery succeeds after about 1,300 observations and in less than two minutes.
Similarly, we extract the private ECDSA key from a hardware TPM manufactured by STMicroelectronics, which is certified at Common Criteria (CC) EAL 4+,
after fewer than 40,000 observations.
We further highlight the impact of these vulnerabilities by 
demonstrating a remote attack against a StrongSwan IPsec VPN that uses a TPM to generate the digital signatures for authentication.
In this attack, the remote client recovers the server's private authentication key by timing only 45,000 authentication handshakes via a network connection.

The vulnerabilities we have uncovered emphasize the difficulty of correctly implementing known constant-time techniques,
and show the importance of evolutionary testing and transparent evaluation of cryptographic implementations.
Even certified devices that claim resistance against attacks require additional scrutiny by the community and industry, 
as we learn more about these attacks.

\end{abstract}

\section{Introduction}\label{sec:intro}
 
Hardware support for trusted computing has been proposed based on trusted execution environments (TEE) and secure elements such as the Trusted Platform Module (TPM)~\cite{mitchell2005trusted}.
Computer manufacturers have been deploying TPMs on desktop workstations, laptops, and servers for over a decade.
With a TPM device attached to the computer, the root of trust can be executed in a separate hardened cryptographic core, which prevents even a fully compromised OS from revealing credentials or keys to adversaries.
TPM 2.0, the latest standard, is deployed in almost all modern computers and is required by some core security services~\cite{howWindowsTPM}.
TPM 2.0 supports multiple signature schemes based on elliptic curves~\cite{tpm2LibSpec}.

TPMs were originally designed as separate hardware modules, but new demands have resulted in software-based implementations.
The physical separation of the TPM from the CPU is an asset for protection against system-leval adversaries~\cite{bajikar2002trusted}.
However, its lightweight design and low-bandwidth bus connection prevents the TPM from being used as a secure cryptographic co-processor for high-throughput applications.
TEE technologies such as ARM TrustZone~\cite{arm2009security} are a more recent approach to bringing trusted execution right into the CPU, at minimal performance loss.
Firmware TPMs (\fTPM) can run entirely in software within a TEE like ARM Trustzone~\cite{raj2016ftpm}.
In a cloud environment, a software-virtualized TPM device will be executed within the trust boundary of the hypervisor~\cite{perez2006vtpm,googlevtpm,azurevtpm}.
In this case, user applications still benefit from the defense against attacks on the guest OS.
Virtual TPMs may or may not rely on a physically present TPM hardware.
Intel Platform Trust Technology (PTT), introduced in Haswell processors, is based on \fTPM\ and follows a hybrid hardware/software approach to implement the TPM 2.0 standard, as discussed in~\autoref{sec:platformsec}.
By enabling Intel PTT, computer manufacturers do not need to deploy dedicated TPM hardware.

Side-channel attacks are a potential attack vector for secure elements like TPMs.
These attacks exploit the unregulated physical behavior of a computing device to leak secrets.
Processing cryptographic keys may expose secret-dependent signal patterns through physical phenomena such as power consumption, electromagnetic emanations, or timing behavior~\cite{quisquater2001electromagnetic,mangard2008power,brumley2005remote}. 
A passive adversary who observes such signals can reconstruct cryptographic keys and break the confidentiality and authenticity of a computing system~\cite{messerges2002examining,den2002dpa}.
The TPM, as defined by the Trusted Computed Group (TCG), attempts to mitigate the threat of physical attacks through a rigorous and lengthy evaluation and certification process.
Most physical TPM chips have been certified according to Common Criteria, which involves evaluation through certified testing labs.
Tests are conducted according to protection profiles. For TPM, a specific TCG protection profile exists, which requires the TPM to be protected against side-channel attacks, 
including timing attacks~\cite[p.~23]{TCGPP}.

TPMs have previously suffered from vulnerabilities due to weak key generation~\cite{nemec2017return}.
However, it is widely believed that the execution of cryptographic algorithms is secure even against system adversaries.
Indeed, TPM devices are expected to provide a more reliable root of trust than the OS by keeping cryptographic keys secure.
Contrary to this belief, we show that these implementations can be vulnerable to remote attacks.
These attacks not only reveal cryptographic keys, but also render modern applications using the TPM less secure than without the TPM.



\subsection{Our Contribution}
In this work, we perform a black-box timing analysis of TPM devices.
Our analysis reveals that elliptic curve signature operations on TPMs from various manufacturers are vulnerable to timing leakage that leads to recovery of the private signing key.
We show that this leakage is significant enough to be exploited remotely by a network adversary.
In summary, our contribution includes:

\begin{itemize}
    \item An analysis tool that can accurately measure the execution time of TPM operations on commodity computers.
    Our developed tool supports analysis of command response buffer (CRB) and TPM Interface Specification (TIS) communication interfaces.
    \item The discovery of previously unknown vulnerabilities in TPM implementations of ECDSA and ECSchnorr signature schemes, and the pairing-friendly BN-256 curve used by the ECDAA signature scheme.
    These elliptic curve signature schemes are supported by the TPM 2.0 standard.  
    We apply lattice-based techniques to recover private keys from these side-channel vulnerabilities. 
    \item A remote attack that breaks the authentication of a VPN server that uses Intel fTPM to store the private certificate key and to sign the authentication message.
    We demonstrate the efficacy of our attack against the strongSwan IPsec-based VPN Solution that uses the TPM device to sign authentication messages.
\end{itemize}
Our study shows that these vulnerabilities exist in devices that have been validated based on FIPS 140-2 Level 2 and Common Criteria (CC) EAL 4+, 
which is the highest internationally accepted assurance level in CC, in a protection profile that explicitly includes timing side channels.

\subsection{Experimental Setup}
We tested Intel fTPM on multiple computers running Intel Management Engine (ME), 
and we demonstrate key recovery attacks on these machines. 
We also tested multiple machines manufactured with dedicated TPM hardware, as discussed in~\autoref{sec:analysis}. 
All the machines run Ubuntu 16.04 with kernel 4.15.0-43-generic.
We used the \textit{tpm2-tools}\footnote{https://github.com/tpm2-software/tpm2-tools \texttt{commit c66e4f0}} and \textit{tpm2-tss}\footnote{https://github.com/tpm2-software/tpm2-tss \texttt{commit 443455b}} software packages and the default TPM kernel device driver to interact with the TPM device.
Our analysis tool takes advantage of a custom Linux loadable kernel module (LKM).

The remote attacks are demonstrated on a simple local area network (LAN) with the attacker and victim workstation connected through a 1 Gbps switch manufactured by Netgear.

\subsection{Coordinated Disclosure}
We informed the Intel Product Security Incident Response Team (iPSIRT) of our findings regarding Intel \fTPM\ on February 1, 2019.
Intel acknowledged receipt on the same day, and responded that an outdated version of Intel IPP has been used in the Intel \fTPM\ on February 12, 2019.
Intel assigned CVE-2019-11090 and awarded us separately for three vulnerabilities.
They issued a firmware update for Intel Management Engine (ME) including patches to address this issue on November 12, 2019.

We informed STMicroelectronics of our findings regarding the TPM chip flaw on May 15, 2019.
They acknowledged receipt on May 17, 2019.
We shared our tools and techniques with STMicroelectronics.  
They assigned CVE-2019-16863 and provided us an updated version of their TPM product for verification.
We tested the updated hardware and confirmed that it is resistant to our attacks on September 12, 2019.

\section{Background}
\subsection{Trusted Platform Module}
\begin{figure}[t!]
    \centering
\includegraphics[width=0.90\linewidth, keepaspectratio]{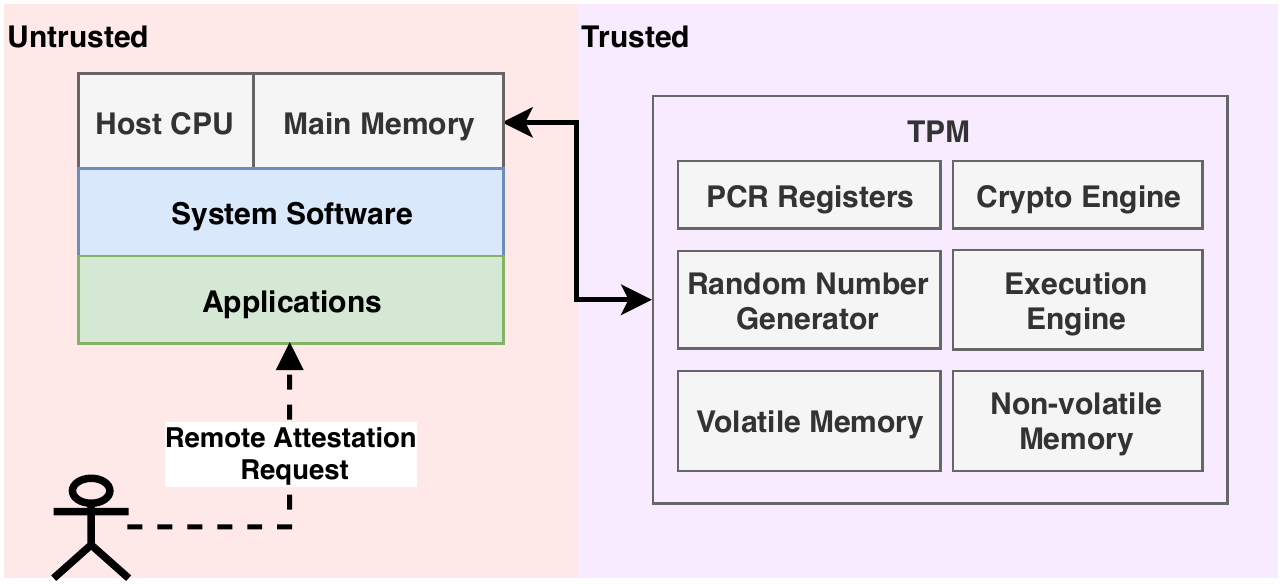} 
\caption{The trusted components of a TPM include the PCR registers, crypto engine, and random number generator. Other hardware components, system software, and applications are considered untrusted.}
    \label{fig:tpm}
\end{figure}

TPMs are secure elements which are typically dedicated physical chips with Common Criteria certification at EAL 4 and higher, and thus provide a very high level of security assurance for the services they offer~\cite{Challener2011}.
As shown in~\autoref{fig:tpm}, the TPM, including components like cryptographic engines, forms the root of trust.
On a commodity computer, the host processor is connected to the TPM via a standard communication interface~\cite{tcg2005client}.
For trusted execution of cryptographic protocols, applications can request that the OS interact with the TPM device and use various cryptographic engines that support hash functions, encryption, and digital signatures.
The TPM also contains non-volatile memory for secure storage of cryptographic parameters and configurations.
As discussed in~\autoref{sec:strongswan}, for instance, a Virtual Private Network (VPN) application can use the TPM to securely store authentication keys and to perform authentication without direct access to the private key.
TPM also supports remote attestation, in which the TPM will generate a signature using an attestation key which is normally derived from the device endorsement key.
The endorsement key is programmed into the TPM during manufacturing. 
Later on, the signature and the public attestation key can be used by a remote party to attest to the integrity of the system, and 
the public endorsement key can be used to verify the integrity of the TPM itself.

{\bf Attacks on TPM:}
The traditional communication interface between dedicated TPM hardware and the CPU is the Low Pin Count (LPC) bus, which has been shown to be vulnerable to passive eavesdropping~\cite{kursawe2005analyzing}.
There exist attacks to compromise the PCRs based on short-circuiting the LPC pins~\cite{sparks2007security,kauer2007oslo}, software-based attacks on the BIOS and bootloader~\cite{kauer2007oslo,butterworth2013bios}, 
and attacks exploiting vulnerabilities related to the TPM power management~\cite{han2018bad}. 
Nemec at al.\ developed the ``Return of Coppersmith's Attack'' (ROCA), which demonstrated passive RSA key recovery from the public key resulting from the special structure of primes generated on TPM devices manufactured by Infineon~\cite{nemec2017return}.
The remote timing attacks that we demonstrate are orthogonal to the key generation issues responsible for ROCA. 
As originally suggested by Spark. et al.~\cite{sparks2007security}, we demonstrate a class of remote timing attack against TPM devices that are deployed within hundreds of thousands of desktop/laptop computers.

\subsection{Intel Management Engine}
\label{sec:platformsec}
The Intel management engine (ME) provides hardware support for various technologies such as Intel Active Management Technology (AMT), Intel SGX Enhanced Privacy ID (EPID) provisioning and attestation, and platform trust technology (PTT)~\cite{ververis2010security}.
Intel ME is implemented as an embedded coprocessor that is integrated into all Intel chipsets.
This coprocessor runs modular firmware on a tiny microcontroller.
Since the Skylake generation, Intel has used the MINIX3 OS running on a 32-bit Quark x86 microcontroller\footnote{Quark microcontrollers have a working frequency of 32\,MHz~\cite{intelQuark}.}.
These firmware modules, and in particular the cryptographic module, provide commonly used functions for a variety of services.
Previous reverse-engineering efforts have uncovered some of the secrets of the Intel ME implementation~\cite{skochinsky2014intel}, 
as well as classical software flaws and vulnerabilities related to the JTAG that can be abused to compromise Intel ME~\cite{ermolov2017hack,Ermolov2017JTAG,ermolov2019hack}. 

Intel PTT, which is essentially a firmware-based TPM, has been implemented as a module that runs on top of the Intel Management Engine (ME).
Intel PTT executes on a general purpose microcontroller, but since it executes independently from the host processor components, 
it resembles a more secure hybrid approach than the original Intel \fTPM~\cite{raj2016ftpm}, which executes on a TEE on the same core.
The exact implementation of the cryptographic functions that are shared by Intel PTT, EPID, and other cryptographically relevant services is not publicly available.

\subsection{Elliptic Curve Digital Signatures} \label{sec:ecc}
The Elliptic Curve Digital Signature Algorithm (ECDSA)~\cite{johnson2001elliptic} is an elliptic curve variant of the Digital Signature Algorithm (DSA)~\cite{gallagher2013digital} in which 
the prime subgroup in DSA is replaced by a group of points on an elliptic curve over a finite field.
The ECDSA key generation process starts with the selection of an elliptic curve, specified by the curve parameters and the base field $\F_q$ over which the curve is defined, and 
a base point $P \in E$ of cryptographically large order $n$ in the group operation. 

\medskip
\noindent{\bf ECDSA Key Generation:}
\setlist{nolistsep}
\begin{enumerate}
\item Randomly choose a private key $d \in  \Z_n^*$.
\item Compute the curve point $Q=dP \in \mathcal{E}$.
\end{enumerate}
The private, public key pair is $(d, Q)$.

\medskip
\noindent
{\bf ECDSA Signing:}
To sign a message $m\in\{0,1\}^*$
\begin{enumerate}
\item Choose a nonce/ephemeral key $k \in  \Z_n^*$.
\item Compute the curve point $kQ$, and compute the $x$ coordinate $r = (kQ)_x$.
\item Compute $s = k^{-1}(H(m) + dr) \bmod{n}$, where $H(.)$ represents a cryptographic hash function such as SHA-256.
\end{enumerate}
The signature pair is $(r,s)$.

The Schnorr digital signature scheme~\cite{schnorr1991efficient} has been similarly extended to support elliptic curves.
Among multiple different standards for Elliptic Curve Schnorr (ECSchnorr), the TPM 2.0 is based on the ISO/IEC 14888-3 standard.

The key generation for ECSchnorr is similar to ECDSA. The signing algorithm is defined as the following: 

\medskip
\noindent
{\bf ECSchnorr Signing:}
To sign a message $m\in\{0,1\}^*$,
\begin{enumerate}
\item Choose an ephemeral key $k \in  \Z_n^*$.
\item Compute the elliptic curve point $kQ$ and compute the $x$ coordinate $xR = (kQ)_x$.
\item Compute $r = H(xR\,||\,m) \bmod{n}$.
\item Compute $s = (k + dr) \bmod{n}$.
\end{enumerate}
The signature pair is $(r,s)$. 

In practice, elliptic curve signature schemes are implemented for a small set of standard curves, which have been vetted for security.
The targeted elliptic curves that we will discuss in this paper are the \texttt{p-256}~\cite{gallagher2013digital} and \texttt{bn-256}~\cite{barreto2005pairing} curves, as supported by TPM 2.0.
\texttt{bn-256} can optionally be used with ECDSA and ECSchnorr schemes, but it is essential for the elliptic-curve direct anonymous attestation (ECDAA) scheme, 
since ECDAA requires a pairing-friendly curve like \texttt{bn-256}. 
Since it is not relevant to our attack, we omit discussion of ECDAA and signature verification.

\subsection{Lattice and Timing Attacks}
{\bf The Hidden Number Problem:} 
Boneh and Venkatesan~\cite{boneh1996hardness} formulated the hidden number problem (HNP) as the following:
Let $\alpha \in \mathbb{Z}_p^*$ be an integer that is to remain secret.
In the hidden number problem, one is given a prime $p$, several uniformly and independently randomly chosen integers $t_i$ in $\mathbb{Z}^*_p$, 
and also integers $u_i$ that represent the $l$ most significant bits of $\alpha t_i \bmod p$.
The $t_i$ and $u_i$ satisfy the property $| \alpha t_{i} - u_{i} | < p / 2^l$.
Boneh and Venkatesan showed how to recover the secret integer $\alpha$ in polynomial time using lattice-based algorithms with probability greater than $1/2$, 
if the attacker learns enough samples from the $l$ most significant bits of $\alpha t_i\bmod p$. 

{\bf Lattice Attacks:} 
Researchers have applied lattice-based algorithms for the HNP to attack the DSA and ECDSA signing algorithms with partially known nonces~\cite{howgrave2001lattice,romer2001information,nguyen2002insecurity,nguyen2003insecurity}.
As a direct consequence, implementation of these signature algorithms in standard cryptographic libraries have been shown to be vulnerable when the implementation leaks partial information about the secret nonce through side channels~\cite{fan2016attacking,pereida2016make,benger2014ooh,ryan2019return}.
Lattice attacks can also solve similar HNP instances to recover private keys for other signature schemes such as EPID in the presence of side channel vulnerabilities~\cite{dall2018cachequote}.
Ronen et al.~\cite{ronen20199} connected padding oracle attacks to the HNP.
While there exist other variants of the HNP, such as the modular inversion hidden number problem~\cite{boneh2001modular} and 
the extended hidden number problem~\cite{hlavavc2006extended},
our attack is based on the original HNP where the attacker learns information about the most significant bits of the nonce. 
A second family of algorithms for solving the HNP is based on Fourier analysis. 
Bleichenbacher's algorithm~\cite{bleichenbacher2005experiments} was the first to make this connection. 
Bleichenbacher's Fourier analysis techniques can be augmented with lattice reduction for the first stage of the attack, as shown by De Mulder et al.~\cite{de2014using}. 
Bleichenbacher's original algorithm is targeted at a scenario where only a very small amounts of information is leaked by each signature, 
and the attacker can query for a very large number of signatres; the De Mulder variant requires fewer signatures, but in this setting the above lattice techniques are more efficient.
We use lattice attacks because they are more efficient for the amount of side-channel information we obtain.

{\bf Timing Attacks:}
Kocher showed that secret-dependent timing behavior of cryptographic implementations can be used to recover secret keys~\cite{kocher1996timing}.
Since then, constant-time operation, or at least secret-independent execution time, has become a common requirement for cryptographic implementations.
For example, the Common Criteria evaluation of cryptographic modules, which is common for standalone TPMs, includes testing for timing leakage.
Brumley et al.~showed that remote timing attacks can be feasible across networks by mounting an attack against RSA decryption as it was implemented in OpenSSL~\cite{brumley2005remote}.
Similarly, the OpenSSL ECDSA implementation was vulnerable to remote timing attacks~\cite{brumley2011remote}.
In the latter work, they also showed how lattice attacks can be used to recover private keys based on the nonce information.
However, the practicality of such attacks has been questioned in the real world~\cite{wong2015timing} due to noise and low timing resolution.

In comparison, we show that such timing attacks have a greater impact on TPMs, 
because of the high-resolution timing information and their specific threat model of a system-level attacker. 
Timing side channels have also been used to attack the implementation of cryptographic protocols.
For example, both the Lucky 13 attack~\cite{al2013lucky} and Bleichenbacher's RSA padding oracle attack~\cite{meyer2014revisiting} exploit remote timing.

\section{Timing Attack and Leaky Nonces} \label{sec:analysis}
Our timing attacks have three main phases: 

{\bf Phase 1:}
The attacker generates signature pairs and timing information and uses this information to profile a given implementation.  The timing oracle can be based on a remote source, for example the network round-trip time, or precise local source, as discussed in~\autoref{sec:timing}.  In this pre-attack profile stage, the attacker knows the secret keys and can use this to recover the nonces, and thus has perfect knowledge of the correlation between timing and partial information about the secret nonce $k$ that is leaked through this timing oracle.
As explained in~\autoref{sec:vulns}, in our case this bias is related to the number of leading zero bits (LZBs) in the nonce, which is revealed by the timing oracle. 
For the vulnerable TPM implementations in this paper, signing a message with a nonce that has more leading zero bits is expected to take less time.

{\bf Phase 2:} 
To mount a live attack, the attacker has access to a secret-related timing oracle as above and collects a list of signature pairs and timing information from a vulnerable TPM implementation. 
The attacker uses the signature timing information obtained during the profiling phase to filter out signatures and only keep the signature pairs $(r_i$, $s_i)$ that have a specific bias in the nonce $k_i$.

{\bf Phase 3:} 
The attacker applies lattice-based cryptanalysis to recover the private key $d$ from a list of filtered signatures with biased nonces $k_i$. 
In the noisier cases, e.g. with timings collected remotely over the network, filtering may not work perfectly and the lattice attack may fail. 
In these cases, the attacker can randomly chose subsets of filtered signatures, and repeatedly run the lattice attack with the hope of leaving the noisy samples out.

This section describes our custom timing analysis tool, and shows how a privileged adversary can exploit the OS kernel to perform accurate timing measurement of the TPM, 
and thus discover and exploit timing vulnerabilities in cryptographic implementations running inside the TPM. 
We then report the vulnerabilities we discovered related to elliptic curve digital signatures. 
Later, in~\autoref{sec:actualattacks}, we combine the knowledge of these vulnerabilities with the lattice-based cryptanalysis discussed in~\autoref{sec:latticeattack}
to demonstrate end-to-end key recovery attacks under various practical threat models\footnote{The source code for our timing analysis tool, lattice attack scripts, and a subset of the data set is available at \href{https://github.com/VernamLab/TPM-Fail}{github.com/VernamLab/TPM-Fail.}}.

\subsection{Precise Timing Measurement} ~\label{sec:timing}
The TPM device runs at a much lower frequency than the host processor, as it is generally implemented based on a power-constrained platform such as an embedded microcontroller. 
A modern Intel core processor's cycle count can be used as a high-precision time reference to measure the execution time of an operation inside the TPM device.
In order to perform this measurement on the host processor entirely from software while minimizing noise, 
we need to make sure that we can read the processor's cycle count right before the TPM device starts executing a security-critical function, and right after the execution is completed.

The Linux kernel supports device drivers to interact with the TPM that support various common communication standards.
Our examination of the TPM kernel stack and different TPM 2.0 devices on commodity computers suggests that Intel \fTPM\ uses the command response buffer (CRB)~\cite{tcgCRB}, and dedicated hardware TPM devices use the TPM Interface Specification (TIS)~\cite{tcg2005client} to communicate with the host processor.
The Linux TPM device driver implements a push mode of communication with these interfaces, where the OS sends the user's request to the device, and checks in a loop whether the operation has been completed by the device or not.
As soon as the completed status is detected, the OS reads the response buffer and returns the results to the user. 
The status check for this operation initially waits for 20 milliseconds to perform another status check, 
and it doubles the wait time every time the device is in a pending state.

This push mode of communication makes timing measurement of TPM operations from user space less efficient and prone to noise.
To mitigate the noise, we initially develop a kernel driver that installs hooks into the CRB and TIS interfaces to modify the described behavior, 
and measure the timing of TPM devices as accurately as possible. 
Later, we move to more realistic settings, i.e. noisy user level access without root privileges, then to settings where the TPM is accessed remotely over the network.

\subsubsection{CRB Timing Measurement}
CRB supports a \texttt{control area} structure to interface with the host processor. 
The \texttt{control area}, as shown in~\autoref{tab:crb}, is defined as a memory mapped IO (MMIO) on the Linux OS in which 
the TPM drivers communicate with the device by reading from or writing to this data structure.
We install a hook on the \texttt{crb\_send} procedure that is responsible for sending a TPM command to the device over the CRB interface.
By default, the driver sets the \texttt{Start} field in the control area after preparing the command size and address of the command buffer to trigger the execution of the command by the device.
Later on, the device will clear this bit when the command is completed.
~\autoref{lst:crb} shows the modification of \texttt{crb\_send}, in which the \texttt{Start} field is checked in a tight loop after trigger.
As a result, the \texttt{crb\_send} will only return upon completion of the command, and cycle counts are measured as close to the device interface as possible.

\begin{table}[t!]
\centering
    \caption{The CRB control area: The CRB interface does not prescribe a specific access pattern to the fields of the Control Area. 
    The \texttt{Start} and \texttt{Status} fields are used to start a TPM command and check the status of the device, respectively.}
    \label{tab:crb}
\begin{tabular}{p{2.13cm}p{0.75cm}p{4.3cm}} \hline
    \toprule
    \footnotesize{\textbf{Field}} & \footnotesize{\textbf{Offset}} & \footnotesize{\textbf{Description}} \\
    \midrule
    \footnotesize{Request}          &  \footnotesize{00}  &  \footnotesize{Power state transition control}    \\
    \footnotesize{Status}           &  \footnotesize{04}  &  \footnotesize{Status}     \\
    \footnotesize{Cancel}           &  \footnotesize{08}  &  \footnotesize{Abort command processing} \\
    \footnotesize{Start}            &  \footnotesize{0c}  &  \footnotesize{A command is available for processing} \\
    \footnotesize{Interrupt Control}&  \footnotesize{10}  &  \footnotesize{Reserved} \\
    \footnotesize{Command Size}     &  \footnotesize{18}  &  \footnotesize{Size of the Command (CMD) Buffer} \\
    \footnotesize{Command Address}  &  \footnotesize{1c}  &  \footnotesize{Physical address of the CMD Buffer} \\ 
    \footnotesize{Response Size}    &  \footnotesize{24}  &  \footnotesize{Size of the Response (RSP) Buffer} \\
    \footnotesize{Response Address} &  \footnotesize{28}  &  \footnotesize{Physical address of the RSP Buffer} \\
    \bottomrule
\end{tabular}
\end{table}

\begin{lstlisting}[label={lst:crb},caption=CRB Timing Measurement,basicstyle=\footnotesize,captionpos=b, frame=tb]
t = rdtsc();
iowrite32(CRB_START_INVOKE, &g_priv->regs_t->ctrl_start);
while((ioread32(&g_priv->regs_t->ctrl_start) & CRB_START_INVOKE) == CRB_START_INVOKE);
tscrequest[requestcnt++] = rdtsc() - t;
\end{lstlisting}

\subsubsection{TIS Timing Measurement}
Similarly, the TIS driver uses a MMIO region to communicate with the TPM device.
The first byte of this mapped region indicates the status of the device.
To measure accurate timing of the TPM over TIS, we install a hook on the \texttt{tpm\_tcg\_write\_bytes} procedure.
In the modified handler (\autoref{lst:tis}), we check if the write operation issued by the TIS driver stack is related to the trigger for the command execution, \texttt{TPM\_STS\_GO}.
If this is the case, we check the buffer for \texttt{TPM\_STS\_DATA\_AVAIL} status, indicating the completion of the command execution, in a tight loop.
Similar to CRB, the cycle counts are measured close to the device interface. 

\begin{lstlisting}[label={lst:tis},caption=TIS Timing Measurement,basicstyle=\footnotesize,captionpos=b, frame=tb]
    enum tis_status {TPM_STS_GO = 0x20, 
        TPM_STS_DATA_AVAIL = 0x10, ...};
    int tpm_tcg_write_bytes_handler(struct tpm_tis_data *data, 
        u32 addr, u16 len, u8 *value){
      ...
    if(len == 1 && *value == TPM_STS_GO && TPM_STS(data->locality) == addr)  {
    t = rdtsc();
    iowrite8(*value, phy->iobase + addr);
    while(!(ioread8(phy->iobase + addr) & TPM_STS_DATA_AVAIL));
    tscrequest[requestcnt++] = rdtsc() - t;
    } ...
\end{lstlisting}

\begin{table*}[t!]
\centering
    \caption{Tested Platforms with Intel \fTPM\ or dedicated TPM device.}
    \label{tab:machines}
\begin{tabular}{ccccccc} 
\toprule
\footnotesize{\textbf{Machine}}             & \footnotesize{\textbf{CPU}}            & \footnotesize{\textbf{Vendor}}     & \footnotesize{\textbf{TPM}}           & \footnotesize{\textbf{Firmware/Bios}}       & \footnotesize{\textbf{ECDSA}} \scriptsize{(Cycle)} & \footnotesize{\textbf{RSA}} \scriptsize{(Cycle)}\\ 
\midrule
\footnotesize{NUC 8i7HNK}          & \footnotesize{Core i7-8705G}  & \footnotesize{Intel}      & \footnotesize{PTT (fTPM)}    & \footnotesize{NUC BIOS 0053}       & \footnotesize{4.1e8}  &  \footnotesize{7.0e8}\\
\footnotesize{NUC 7i3BNK}          & \footnotesize{Core i3-7100U}  & \footnotesize{Intel}      & \footnotesize{PTT (fTPM)}    & \footnotesize{NUC BIOS 0076}       & \footnotesize{3.2e8}  &  \footnotesize{5.4e8}\\ 
\footnotesize{Asus GL502VM}        & \footnotesize{Core i7-6700HQ} & \footnotesize{Intel}      & \footnotesize{PTT (fTPM)}    & \footnotesize{Latest OEM}          & \footnotesize{3.5e8}  &  \footnotesize{5.9e8}\\ 
\footnotesize{Asus K501UW}         & \footnotesize{Core i7 6500U}  & \footnotesize{Intel}      & \footnotesize{PTT (fTPM)}    & \footnotesize{Latest OEM}          & \footnotesize{3.4e8}  &  \footnotesize{5.8e8}\\ 
\footnotesize{Dell XPS 8920}       & \footnotesize{Core i7-7700}   & \footnotesize{Intel}      & \footnotesize{PTT (fTPM)}    & \footnotesize{Dell BIOS 1.0.4}     & \footnotesize{4.7e8}  &  \footnotesize{8.0e8}\\
\footnotesize{Dell Precision 5510} & \footnotesize{Core i5-6440HQ} & \footnotesize{Nuvoton}    & \footnotesize{rls NPCT}      & \footnotesize{NTC 1.3.2.8}         & \footnotesize{4.9e8}  &  \footnotesize{1.8e9}\\ 
\footnotesize{Lenovo T580}         & \footnotesize{Core i7-8650U}  & \footnotesize{STMicro}    & \footnotesize{ST33TPHF2ESPI} & \footnotesize{STMicro 73.04}       & \footnotesize{8.7e7}  &  \footnotesize{9.2e8}\\
\footnotesize{NUC 7i7DNKE}         & \footnotesize{Core i7-8650U}  & \footnotesize{Infineon}   & \footnotesize{SLB 9670}      & \footnotesize{NUC BIOS 0062}       & \footnotesize{1.4e8} &  \footnotesize{5.1e8}\\ 
\bottomrule
\end{tabular}

\end{table*}

\subsection{Timing Analysis of ECDSA}
We profiled the timing behavior of the ECDSA signature schemes using the NIST-256p curve.
As shown in~\autoref{tab:machines}, we report the average number of CPU cycles to compute the ECDSA signatures for the aforementioned platforms.
This average cycle count for Intel \fTPM\ is different for each configuration due to the CPU's working frequency, but the average execution time is similar in different configurations: 
for example, we observe the highest cycle count on the Core i7-7700 machine, which is a desktop processor with base frequency of 3.60 GHz. 
We can calculate the average execution time for ECDSA on Intel \fTPM\ as $4.7\times 10^{8}\ \mbox{cycles} / 3.6\ GHz = 130 ms$. 
As mentioned in~\autoref{sec:platformsec}, the working frequency of the Intel \fTPM\ device is relatively slow, which facilitates our observation of timing vulnerabilities on such platforms. 
As the numbers for the dedicated hardware TPM chips suggest, there is a significant difference in execution time between different implementations among various manufacturers.

To test the ECDSA signature scheme, we generated a single ECDSA key using the TPM device, and then measured the execution time for ECDSA signature generation on the device.
As mentioned in~\autoref{sec:ecc}, the security of ECDSA signatures depends on the randomly chosen nonce.
The TPM device must use a strong random number generator to generate this nonce independently and randomly for each signing operation to preserve the security of the ECDSA scheme~\cite{nguyen2003insecurity}. 

Our analysis reveals that Intel \fTPM\ and the dedicated TPM manufactured by STMicroelectronics leak information about the secret nonce in elliptic curve signature schemes, 
which can lead to efficient recovery of the private key.
As discussed in~\autoref{sec:findings}, we also observe non-constant-time behavior by the TPM manufactured by Infineon which does not appear to expose an exploitable vulnerability.
From our experimental observations, only the TPM manufactured by Nuvoton exhibits constant-time behavior for ECDSA (\autoref{fig:ecdsa_nuv}).

\subsection{Discovered Vulnerabilities} \label{sec:vulns}

\begin{figure}[t!]
    \centering
\includegraphics[width=0.90\linewidth]{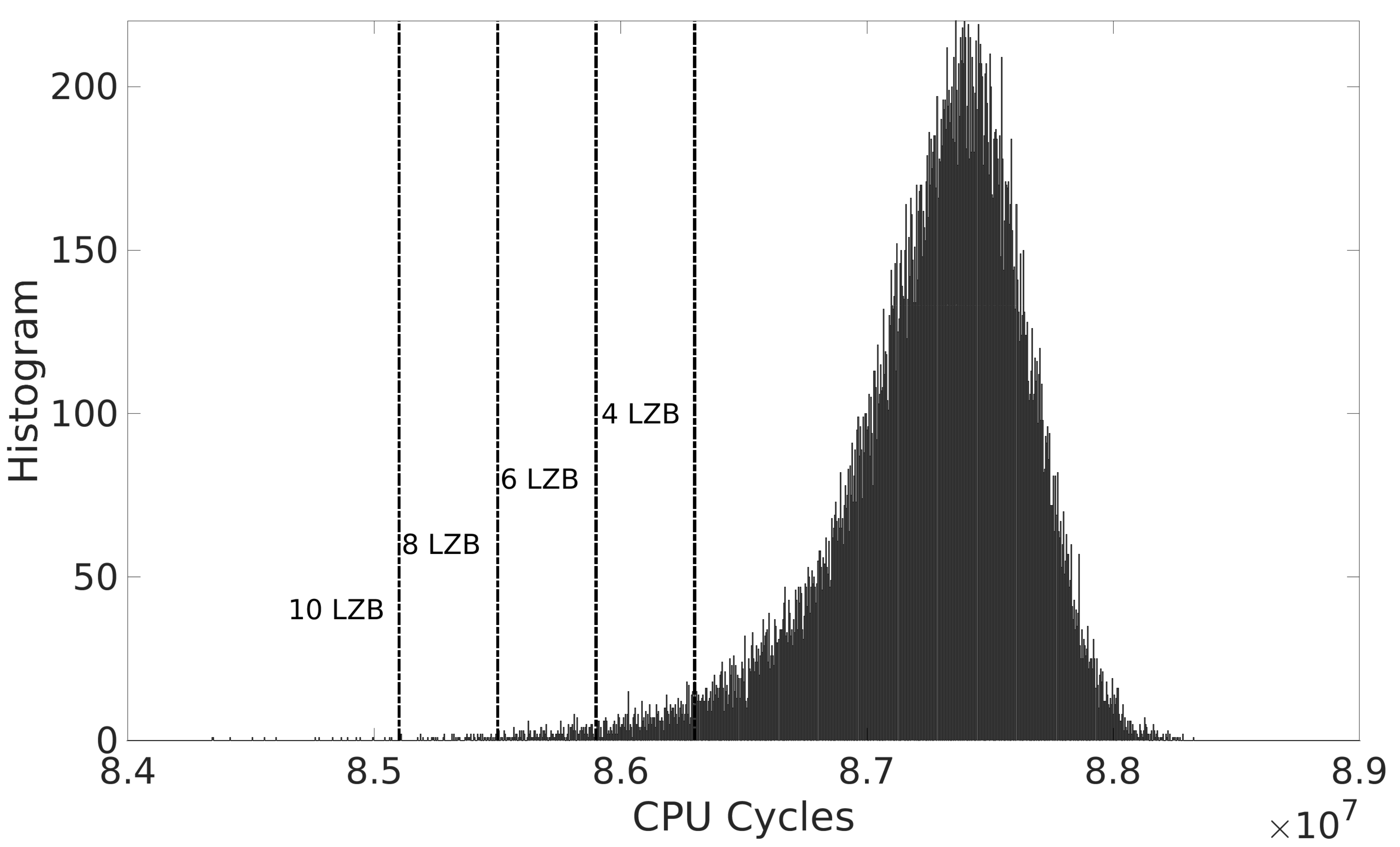} 
\caption{Histogram of ECDSA (NIST-256p) signature generation timings on the STMicroelectronics TPM as measured on a Core i7-8650U machine for 40,000 observations.}
    \label{fig:ecdsa_stm}
\end{figure}

\begin{figure}[t!]
    \centering
\includegraphics[width=0.90\linewidth]{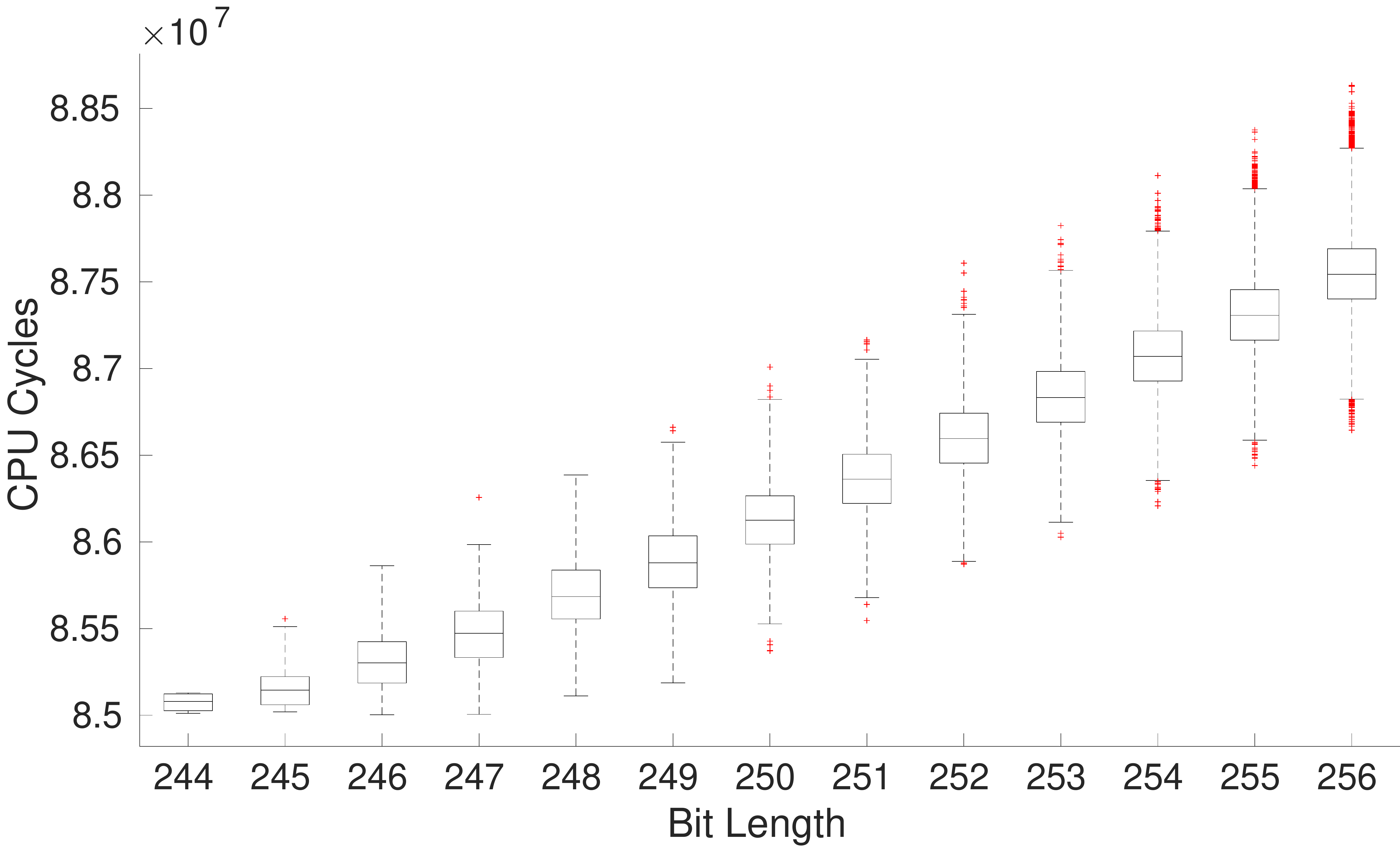} 
\caption{Box plot of ECDSA (NIST-256p) signature generation timings by the bit length of the nonce. We observe a clear linear relationship between the two for the STMicroelectronics TPM. 
Each box plot indicates the median and quartiles of the timing distribution.}
    \label{fig:ecdsa_box_stm}
\end{figure}

\medskip
\noindent
\textbf{STMicroelectronics ECDSA Scalar Multiplication:}
~\autoref{fig:ecdsa_stm} shows an uneven distribution for the STMicroelectronics TPM where there are more leading zero bits (LZBs) on the left side of the distribution. 
We used the private key $d$ to compute each nonce $k_i$ for each profiled signature $(r_i, s_i)$ by computing $k_i = s_i^{-1}(H(m) + dr_i) \bmod{n}$.
~\autoref{fig:ecdsa_box_stm} shows a linear correlation between the execution time and the bit length of nonce. 
This shows that for each additional zero bit, the cycle count differs by an average of $2\times 10^{5}$ cycles. 
This leakage pattern suggests a bit-by-bit scalar point multiplication implementation that skips the computation for the most significant zero bits of the nonce. 
As a result, nonces with more leading zero bits are computed faster.

\medskip
\noindent
\textbf{Intel \fTPM\ ECDSA Scalar Multiplication:}
~\autoref{fig:ecdsa_intel_local} shows three clearly distinguishable peaks centered around 4.70, 4.74, and 4.78.
Scalar multiplication algorithms to compute $r=(kQ)x$ are commonly implemented using a fixed-window algorithm that iterates window by window over the bits of the nonce to compute the product $kQ$ of the scalar $k$ and point $Q$.
In some implementations, the most significant window (MSW) starts at the first non-zero window of most significant bits of the scalar, which may leak the number of leading zero bits of the scalar~\cite{dall2018cachequote}.
With respect to the observed leakage behavior (\autoref{fig:ecdsa_intel_local}), we expect that:

\begin{figure}[t!]
    \centering
\includegraphics[width=0.90\linewidth]{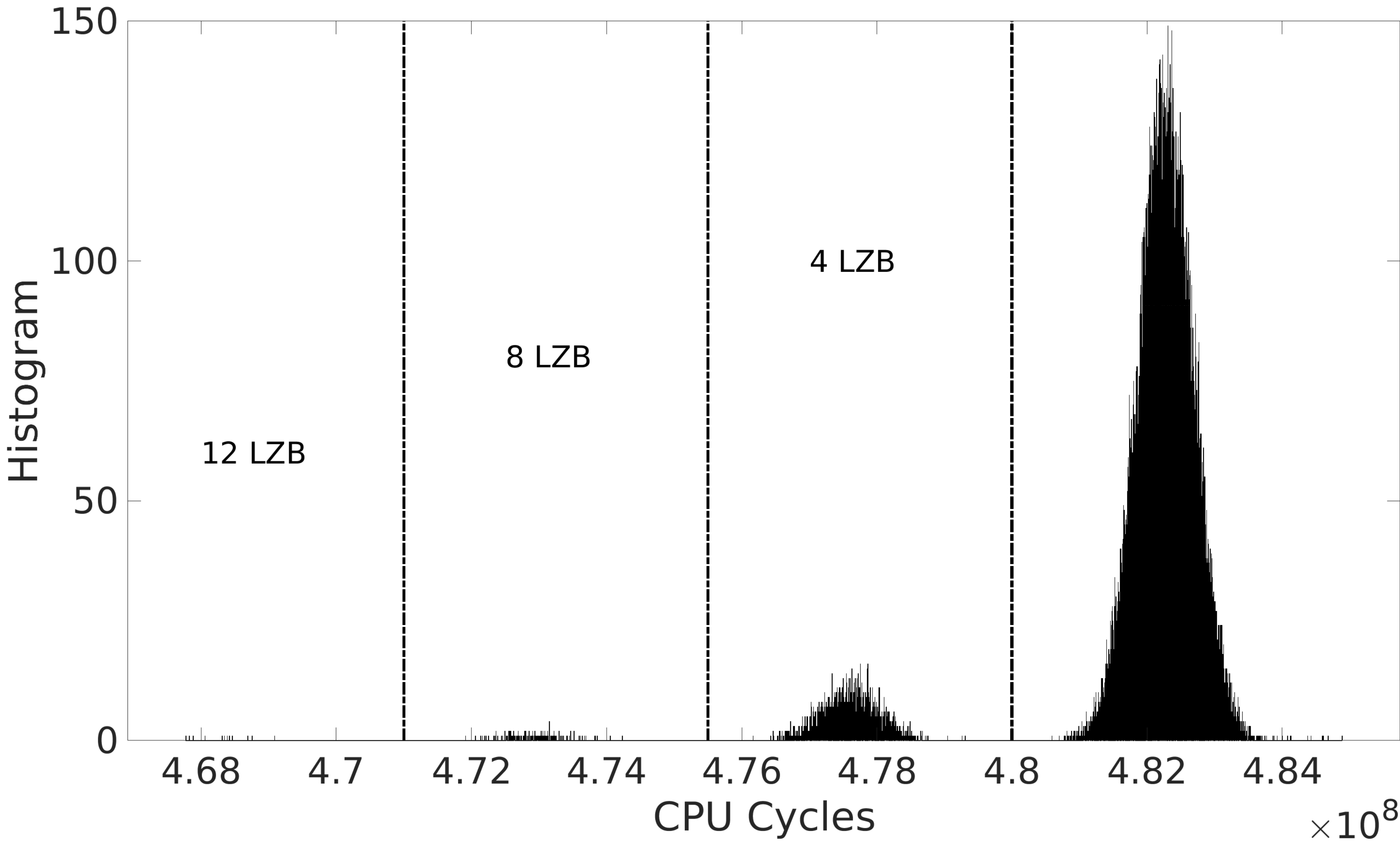} 
\caption{Histogram of ECDSA (NIST-256p) signature generation timings on Intel \fTPM\ as measured on a Core i7-7700 machine for 40K observations.}
    \label{fig:ecdsa_intel_local}
\end{figure}

\begin{itemize}
\item The slowest signatures clustered in the rightmost peak represent those with full length $k$, or in other words, those that have a non-zero most significant window. 
\item The faster signatures clustered in the second peak may represent signatures computed using nonces $k_i$ that have a full zero MSW but a non-zero second MSW.
\item The faster signatures clustered in the third peak may represent signatures computed using nonces $k_i$ that have two full zero MSWs.
\item The fastest signatures on the left peak are generated by nonces with three full MSWs of zero bits. 
\end{itemize}

\begin{figure}[t!]
    \centering
\includegraphics[width=0.90\linewidth]{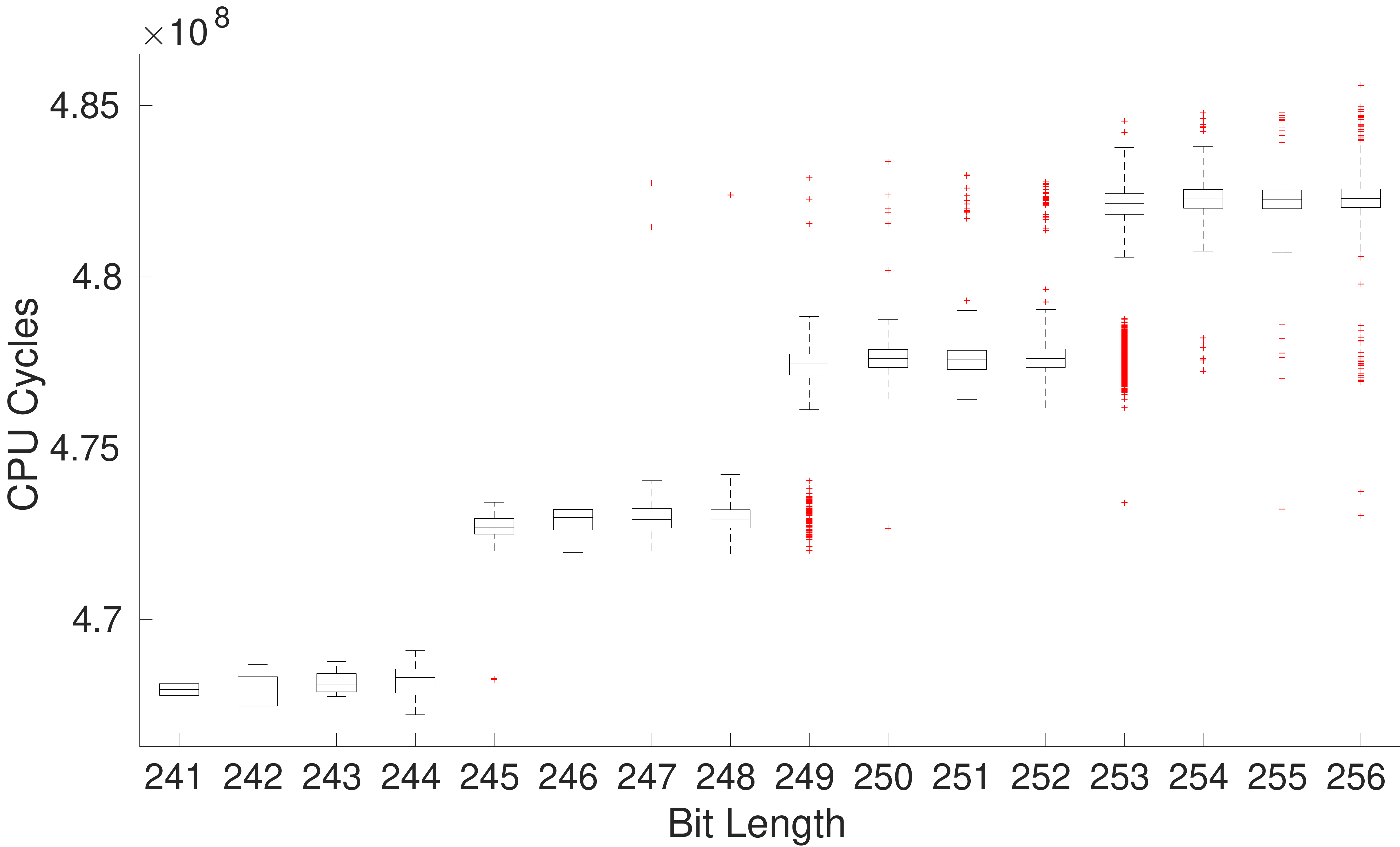} 
\caption{Box plot of ECDSA (NIST-256p) signature generation timings depending on the nonce bit length shows a clear step-wise relationship between the execution time and the bit length of the nonce for Intel fTPM.}
    \label{fig:ecdsa_intel_local_box}
\end{figure}

In addition, the relative sizes of the peaks suggest that the implementation we tested uses a 4-bit fixed window (\autoref{fig:ecdsa_intel_local_box}).
This demonstrates clear leakage of the length of the nonce, which can easily be exploited using a lattice attack.
To summarize, Algorithm~\autoref{alg:fixedmul} matches the observed timing behavior of the scalar multiplication inside the Intel \fTPM.
This observation also aligns with previous vulnerabilities~\cite{wichelmann2018microwalk} which affected earlier versions of Intel IPP cryptography library~\cite{intelIPP}.

\begin{algorithm}[t]
\caption{Fixed Window Scalar Multiplication}
\label{alg:fixedmul}
\begin{algorithmic}[1]
  \State $T \gets (O,P,2P,\ldots,(2w-1)P)$
  \Procedure{MulPoint}{window size $w$, scalar $k$ represented as $(k_{m-1},\ldots,k_0)_{2w}$}
  \State $R \gets T[(k)_{2w}[m-1]]$
  \For{$i \gets m - 2$ \textbf{to} $0$}
    \For{$j \gets 1$ \textbf{to} $w$} 
      \State $R \gets 2R$  
    \EndFor  
  \EndFor
  \State return $R$
\EndProcedure
\end{algorithmic}
\end{algorithm}

\medskip
\noindent
\textbf{Intel \fTPM\ ECSchnorr Scalar Multiplication:}
The ECSchnorr algorithm also uses a secret nonce and scalar multiplication as the first operation of signature generation.
We performed a similar experiment as above, this time using the \texttt{tpm2\_quote} command of the TPM 2.0 device.
\texttt{tpm2\_quote} generates a signature using the configured key, but the signature is computed over the PCR registers rather than an arbitrary message.
The timing observations suggest that ECschnorr executes about 1.4 times faster than ECDSA, which implies an independent implementation, 
but one that is still vulnerable to the same class of timing leakage\footnote{\label{notevuln}The vendor acknowledged this as a separate vulnerability during the bug bounty program. CVE-2019-11090 has been assigned for all issues.} (\autoref{fig:histecschnorr}).

\medskip
\noindent
\textbf{Intel \fTPM\ BN-256 Curve Scalar Multiplication:}
As mentioned earlier, TPM 2.0 also supports the pairing friendly BN-256 curve, which is used as part of the ECDAA signature scheme. 
To simplify our experiment and verify that ECDAA is also vulnerable, we configured ECDSA to operate using the BN-256 curve rather than attacking the ECDAA scheme.
The timing observation of ECDSA is almost doubled by using the BN-256 curve. 
It is also vulnerable, as it leaks the leading zero bits of the secret nonce \footnoteref{notevuln} (\autoref{fig:histbn256}).

\section{Lattice-Based Cryptanalysis} \label{sec:latticeattack}

Now that we have established that our targeted implementations leak information about the nonces used for elliptic curve signatures, 
we show how to use standard lattice techniques to recover the private signing key from this information.

\subsection{Lattice Construction} \label{sec:lattice}
The hidden number problem lattice attacks allow us to recover ECDSA nonces and private keys as long as the nonces are {\em short}.
Since the nonces are uniformly selected from $\Z_n^*$, the $k_i$ will follow an exponentially decreasing distribution of lengths, i.e. half will have a zero in the the most significant bit (MSB), a quarter will have the most significant two bits zero, etc. We will refer to this event as two {\em leading zero bits} or 2 LZBs for short.
Clearly, a randomly selected set of nonces $k_i$ will not be likely to be short, and the lattice attack will not be expected to work. This is where side channels prove invaluable to the attacker. 
 Given some side information that reveals the number of MSBs of $k_i$ that are zero, one can filter out the signatures with short nonces, yielding a set of signatures where the $k_i$ are all short \cite{boneh1996hardness, wong2015timing}.
 This is why having constant-time implementations of DSA and ECDSA schemes is crucial.

To mount an attack on ECDSA, we follow the approach of Howgrave-Graham and Smart~\cite{howgrave2001lattice} and Boneh and Venkatesan~\cite{boneh1996hardness} in reducing ECDSA key recovery to solving the Closest Vector Problem (CVP) in a particular lattice.
We can then follow the strategy outlined by Benger et al.~\cite{benger2014ooh} and embed this lattice into a slightly larger lattice in which the desired vector will appear as a short vector that can be found using standard lattice basis reduction algorithms like {\sf LLL}~\cite{LLL82} or {\sf BKZ}~\cite{BKZ87}. 
Our first step is to define the target lattice from ECDSA signature samples $r_i, s_i$ and $m_i$.
Consider a set of $t$ signature samples $s_i = k_i^{-1}(H(m_i) + dr_i) \bmod{n}$; rearranging slightly, these define a set of linear relations
\[
k_i - s_i^{-1}r_i d - s_i^{-1}H(m_i) \equiv 0 \bmod{n}
\]
where the nonces $k_i$ and the secret key $d$ are unknowns; we thus have $t$ linear equations in $t+1$ unknowns.  Let $A_i= -s_i^{-1}r_i \bmod{n}$ and $B_i=-s_i^{-1}H(m_i) \bmod n$; we thus rewrite our $t$ relations in the form
$k_i+A_i d +B_i = 0 \bmod{n}$.
Let $K$ be an upper bound on the $k_i$.  Now we consider the lattice generated by integer linear combinations of the rows of the following basis matrix
\begin{equation}
M = \begin{bmatrix}
n & & & & & \\
& n & & & & \\
& & \ddots & & & \\
& & & n & & \\ 
A_1 & A_2 & \dots & A_t & K/n & \\
B_1 & B_2 & \dots & B_t &  & K
\end{bmatrix}
\label{eq:hnp}
\end{equation}

The first $t$ columns correspond to each of the $t$ relations we have generated, with the modulus $n$ on the diagonal of each of these columns; the weighting factors of $K/n$ and $K$ in the last two columns have been chosen so that the desired short vector containing the secret key will have coefficients all of approximately the same (small) size, and therefore be more likely to be found than an unbalanced vector.  In particular, this lattice has been constructed so that the vector $v_k = (k_1, k_2, \dots, k_t, K\alpha/n, K)$ is a relatively short vector in this lattice; by construction it is $d$ times the second-to-last row vector of the basis, plus the last vector, with the appropriate integer multiple of $n$ subtracted from each column corresponding to the modular reduction in each of the $t$ relations.  If this vector $v_k$ can be found, the secret key $d$ can be recovered from the second-to-last coefficient of this vector.

Because this target vector $v_k$ is short, we hope that a lattice reduction algorithm like LLL or BKZ might find it, thus revealing the secret key.  The inner workings of these lattice basis reduction algorithms are complex; for the purposes of our attack, we use them as a black box and the only fact that is required is that the LLL algorithm is guaranteed in polynomial time to produce a lattice vector of length $|v| \le 2^{(\dim L -1)/4} (\det L)^{1/\dim L}$; this is an exponential approximation for the shortest vector in the lattice. In practice on random lattices, the LLL algorithm performs somewhat better.  It has been observed to find vectors of length $1.02^{\dim L} (\det L)^{1/\dim L}$~\cite{lllontheaverage}.  For the lattices of relatively small dimension we deal with here, the approximation factor does not play a large role in the analysis, but for large dimensional lattices, the BKZ algorithm achieves a better approximation factor at the cost of an increased running time.
See Boneh and Venkatesan~\cite{boneh1996hardness} and Nguyen and Shparlinksi~\cite{nguyen2002insecurity,nguyen2003insecurity} for a formal analysis and bounds on the effectiveness of this algorithm.

There are two optimizations of this lattice construction that are useful for a practical attack.
The first offers only a minor practical improvement; we can eliminate the variable $d$ by, for example, scaling the first relation by $s_0r_0^{-1}s_i^{-1}r_i$ and subtracting it from the $i^{th}$ equation to obtain $t-1$ linear relations in $t$ unknowns $k_i$, $0 \le i < t$:
\[
k_i-s_0r_0^{-1}s_i^{-1}r_ik_0 - s_i^{-1}H(m_i) + r_0^{-1}s_i^{-1}r_iH(m_i) \equiv 0 \bmod n
\]
This has the effect of reducing the lattice dimension by one.
Otherwise, the lattice construction is the same, except that we replace the $K/n$ scaling factor in the second-to-last row of the basis matrix with a 1.
The second practical optimization is to note that since the $k_i$ are always positive, we can increase the bias by one bit by recentering the nonces around 0.
That is, let $k_i' = k_i - K/2$; if $0 \le k_i \le K$, we now have $-K/2 \le k_i' \le K/2$.  This has the effect of increasing the bias by one bit, which is significant in practice.
We give empirical results applying this attack to our scenario in~\autoref{sec:key-experiments}.

\subsubsection{Modification of the Lattice for ECSchnorr} 
\label{sec:ECSchnorr}
We formulate the problem as in~\autoref{eq:hnp} by writing  
\[
    A_i= -r_0^{-1}r_i \bmod{n} ~~~\mbox{and}~~~
    B_i=s_i^{-1} + s_0r_0^{-1}r_i\bmod n.
\]
At that point, we apply the lattice-based algorithm exactly as in~\autoref{sec:lattice}.

\section{ECDSA Key Recovery on TPMs} \label{sec:actualattacks}
\label{sec:key-experiments}

We put the components of our attacks together to demonstrate end-to-end key recovery attacks in the TPM threat model.
We order the presentation of our attacks from weakest to strongest threat model: 
\textbf{1)} We begin with the strongest adversary, who has system-level privileges with the ability to load Linux kernel modules (LKMs). 
This adversary uses our analysis tool to collect accurate timing measurements. 
\textbf{2)} We reduce the privileges of the adversary to the user-level scenario in which the execution time of the kernel interface can only be measured from user space. 
\textbf{3)} We show how key recovery is still possible with an adversary who can simply measure the network round-trip timings to a remote vicitim.

In all our experiments, we initially programmed the TPM devices with known keys in order to unblind the nonces and facilitate our analysis. 
We have also verified the success of attacks on ST TPM and Intel fTPM using unknown keys generated by each device. 
For this, we used the TPM to internally generate secret keys that remained unknown to us, exported the public key, ran the experiments, and finally verified the recovered secret key using the exported public key.

\subsection{Threat Model I: System-Level Adversary}
In this first attack, we used administrator privileges to collect 40,000 ECDSA signatures and precise timings as shown in the histogram in~\autoref{fig:ecdsa_intel_local}, and filtered the samples to select those with short nonces.
We used the execution time to classify these samples into three conjectured nonce length categories based on the observed 4-bit fixed window: those with four, eight, or twelve most significant bits set to zero. 
We then recovered the nonces and secret keys using the attacks described in~\autoref{sec:lattice}, implemented in {\sf Sage}~8.4~\cite{sagemath} using the {\sf BKZ} algorithm with block size 30 for lattice basis reduction. 
We verified the candidate ECDSA private keys using the public key.
\begin{figure}[t!]
    \centering
\includegraphics[width=0.90\linewidth]{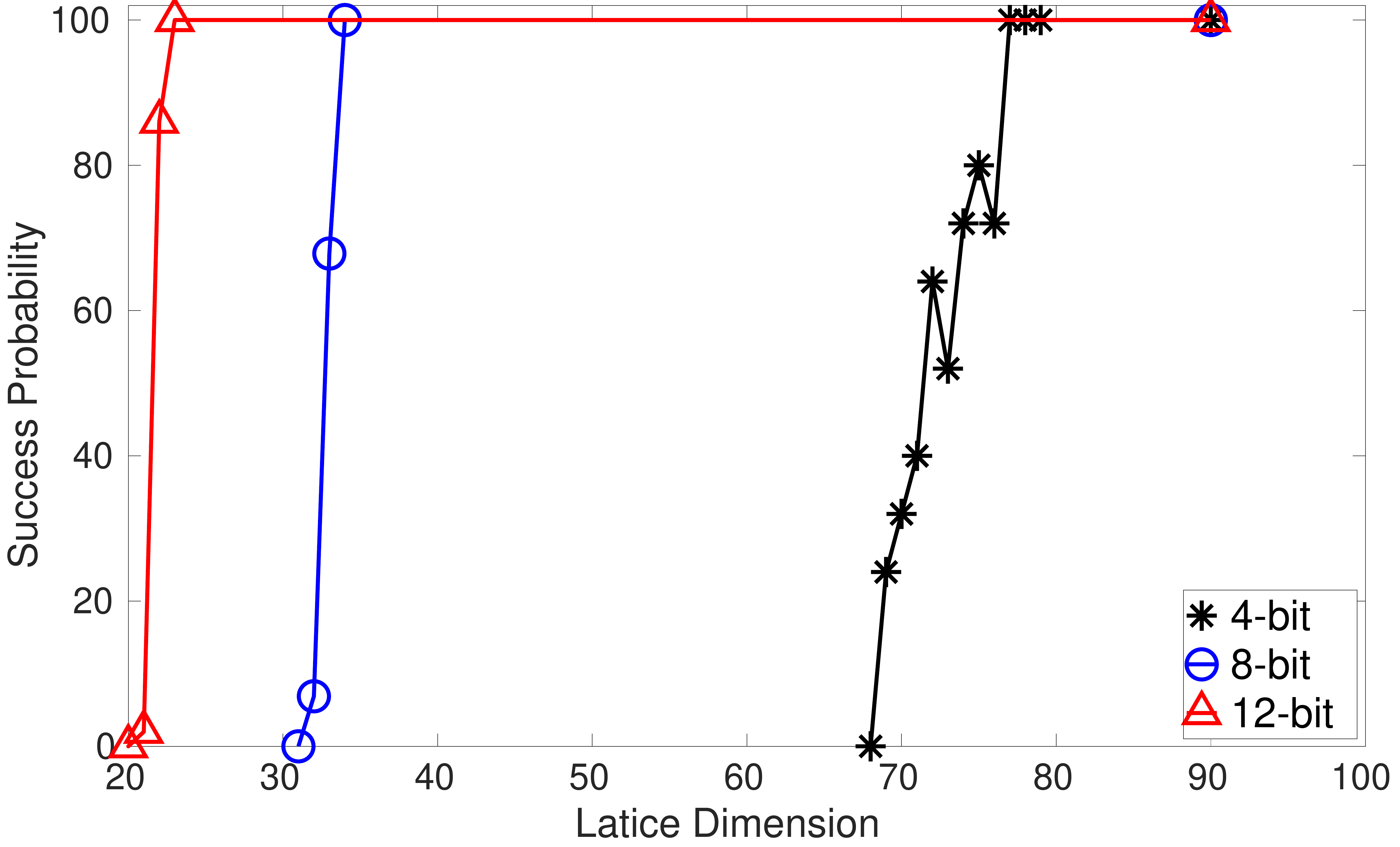} 
\caption{{\bf System Adversary:} Key recovery success probabilities plotted by lattice dimension for 4-, 8-, and 12-bit biases for ECDSA (NIST-256p) with administrator privileges.}
    \label{fig:localroot}
\end{figure}

\autoref{fig:localroot} summarizes the key recovery results for a system-level attacker, 
using samples obtained via simple thresholding with the filter ranges for 12, 8, and 4 LZBs, as shown in~\autoref{fig:ecdsa_intel_local}. 
For example, to recover samples with 4 LZBs, we filtered signatures that took anywhere from $4.75 \times 10^8$ to $4.8 \times 10^8$ cycles to generate.
For the 4-bit bias we need 78 signatures to reach a 92\% key recovery success probability. 
For the 8-bit and 12-bit cases, we can reach 100\% success rate with only 35 and 23 signatures, respectively. 
However, we need to collect more signatures in total in order to generate enough signatures with many LZBs.
The optimal case, with respect to the total number of signature operations, turns out to be using nonces with a 4-bit bias.
Although we need 78 signatures to carry out the attack for the 4-bit bias, since each one occurs with probability of $1/16$, it takes only about 1,248 signing operations to have these samples. 
In our setup on the i7-7700 machine, our collection rate is around 385 signatures/minute.
Therefore, we can collect enough samples in under four minutes.
In the 8-bit case, we need to perform about 8,784 ECDSA signing operations to obtain the 34 suitable signatures necessary for a successful lattice attack.
In total it takes less than 23 minutes to collect 8,784 signatures. 
Once the data is collected, key recovery with lattice reduction takes only 2 to 3 seconds for dimension 30, and about a minute for dimension 70.  The running time of lattice basis reduction can increase quite dramatically for larger lattice dimensions, but the lattice reduction step is not the bottleneck for these attack parameters.

\medskip
\noindent
\textbf{Intel \fTPM\ ECSchnorr Key Recovery:}
We carried out a similar attack against ECSchnorr by modifying the the lattice construction, as described in~\autoref{sec:ECSchnorr}.
We were able to recover the key with 40 samples with 8 LZBs.
A total of 10,240 signatures were required to perform this attack, which can be collected in about 27 minutes.
We also were able to recover the key for the 4-bit case with 65 samples. 
We obtained these 4-bit samples from 1,040 signing operations that took 1.5 minutes to collect.

\medskip
\noindent
\textbf{STMicroelectronics TPM ECDSA Key Recovery:} 
We also tested our approach against the dedicated STMicroelectronics TPM chip (ST33TPHF2ESPI) in the system-level adversary threat model.
This target is Common Criteria certified at EAL4+ for the TPM protection profiles and FIPS 140-2 certified at level 2~\cite{STtpmBrief}.
It is thus certified to be resistant to physical leakage attacks, including timing attacks~\cite{STtpmCCcert}.

We measured the execution times for ECDSA (NIST-256p) signing computations on a Core i7-8650U machine for 115,000 observations.
The machine is equipped with the ST33TPHF2ESPI manufactured by STMicroelectronics.
The administrative privileges allowed us to run our custom driver and collect samples with a high resolution.
Following the vulnerability discussion in~\autoref{sec:vulns}, 
we began by filtering out any data with execution time below $8\times 10^8$ cycles to eliminate noise.
We then sorted the remaining signatures by their execution times.
We were able to recover the ECDSA key after generating 40,000 signatures.
We recovered the key using the fastest 35 signatures and running a lattice attack assuming a bias of 8 most significant zero bits in the nonces.
The required 40,000 samples can be collected in about 80 minutes on this target platform.
We are also able to recover the key from 24 samples by assuming 12 LZBs.
However, this required generating 219,000 total signatures. 

\subsection{Threat Model II: User-Level Adversary}
\begin{figure}[t!]
    \centering
\includegraphics[width=0.90\linewidth]{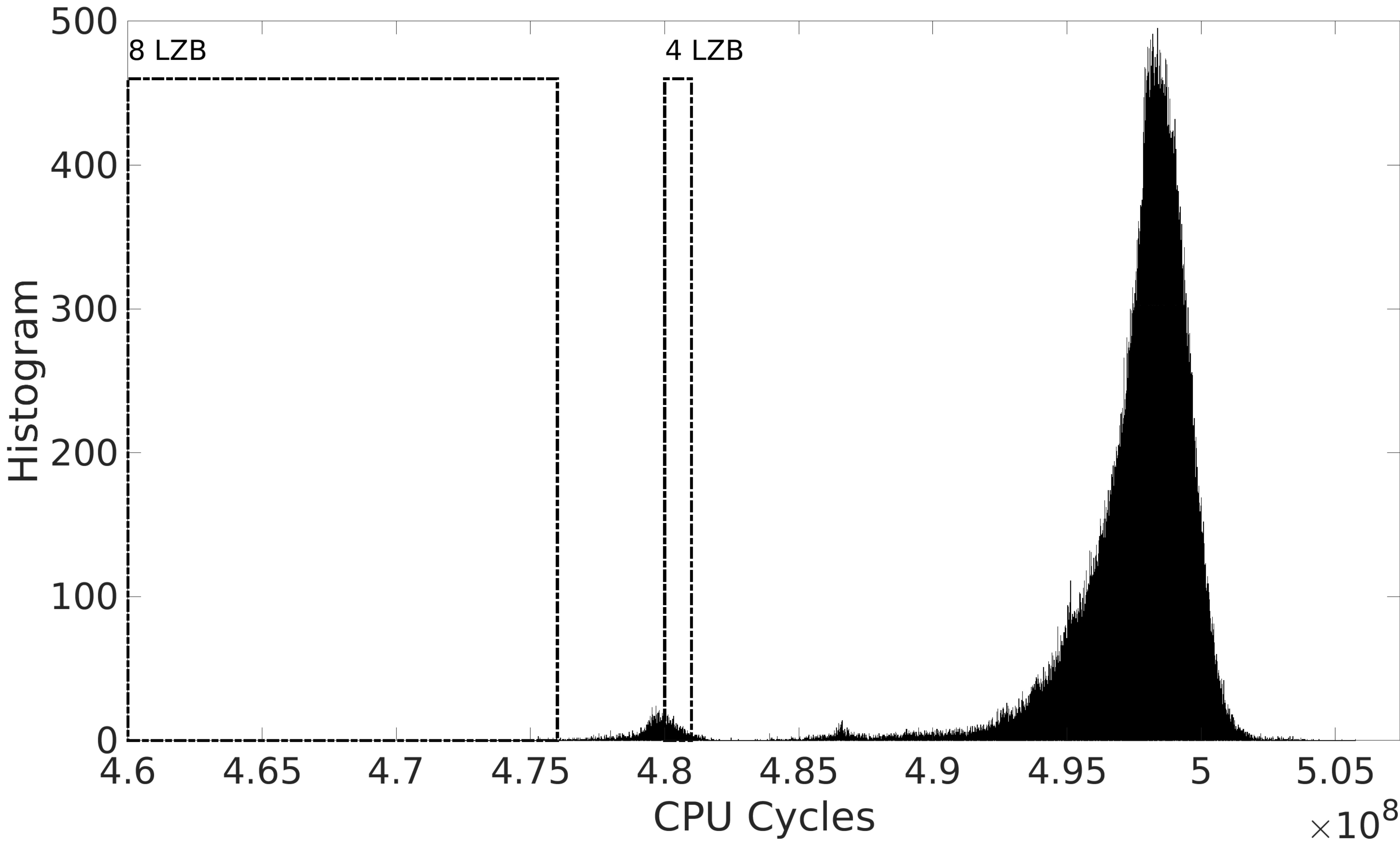}
\caption{{\bf User Adversary:} Histogram of ECDSA (NIST-256p) signature computation times on the Core i7-7700 machine for 40,000 observations. 
The measurements were collected by a user without administrator privileges.}
    \label{fig:ecdsa_intel_user}
\end{figure}

We now move to a less restrictive model, that is, from a system-level adversary to a user-level adversary where only a user API with user-level privileges is provided to perform the signature operations and measure the execution time.
Without the installed kernel measurement tool, we obtain the distribution of signing times shown in~\autoref{fig:ecdsa_intel_user}.
The noise makes it impossible to precisely distinguish the samples according to the number of leading zero bits in the nonces.
However, we observe that we have a biased Gaussian distribution, and by choosing signatures that have a short execution times, we can still recover the ECDSA key. 

\begin{figure}[t!]
    \centering
\includegraphics[width=0.90\linewidth]{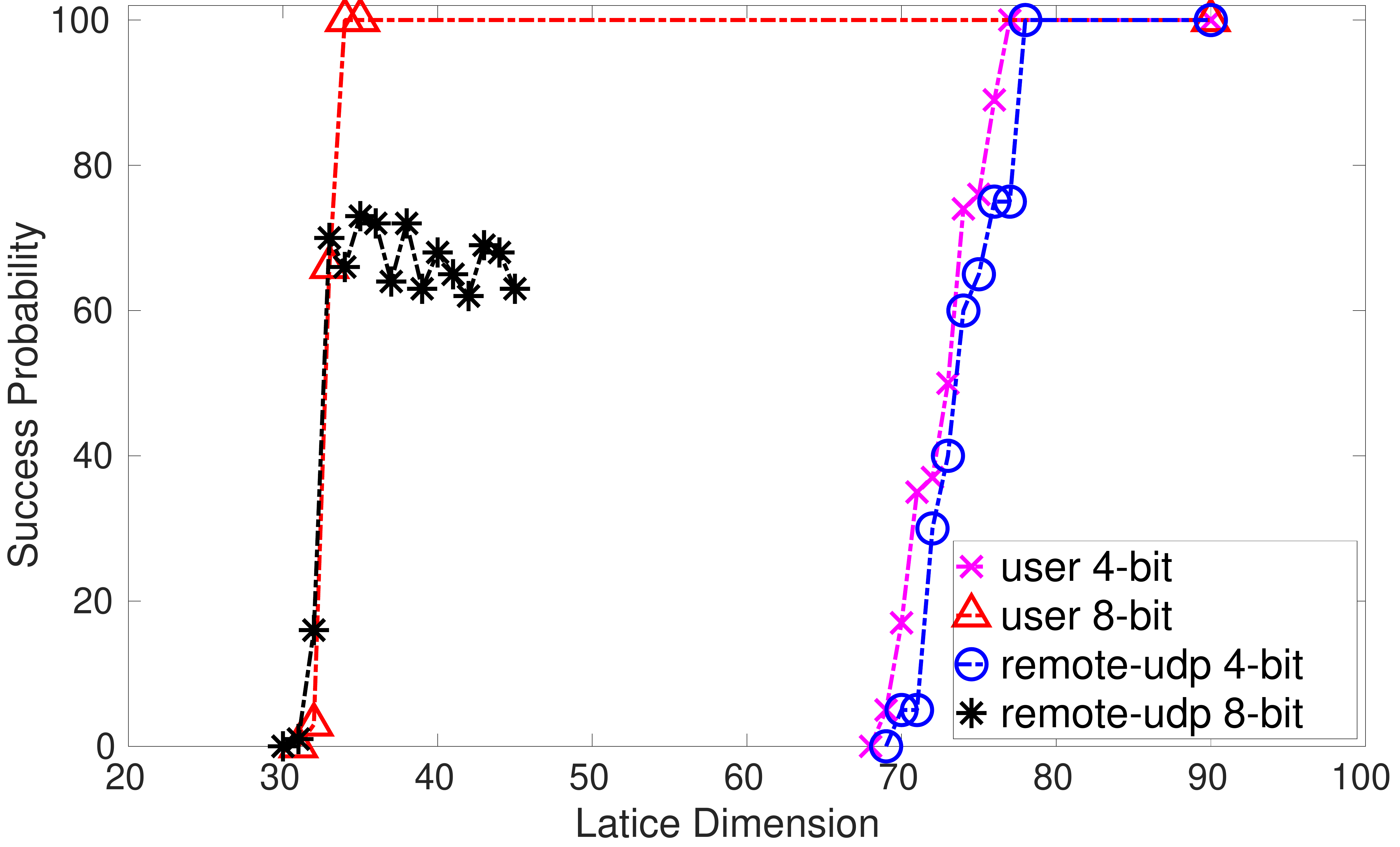}
\caption{{\bf User-Level Adversary and Remote UDP Attack:} 
Key recovery success probabilities by lattice dimension for 4-bit and 8-bit cases for ECDSA (NIST-256p) with timings collected from the user space in one scneario, and over the network from a remote client in another scenario.}
    \label{fig:user_results_udp}
\end{figure}

We start our analysis by noting that in the system-level adversary setting shown in~\autoref{fig:ecdsa_intel_local}, the largest peak is at $4.82\times 10^8$ cycles, 
while in~\autoref{fig:ecdsa_intel_user} the largest peak is around $4.97\times 10^8$. 
This is expected since we incur additional latency by measuring the delay from user space. 
This noise is independent from the bias and therefore we set our filtering thresholds by assuming the entire histogram is shifted by moving the profiling measurements to user space.
We collected a total of 219,000 samples. 
The probability of obtaining a signature sample with 8 LZBs is $1/256$, which means that we expect about $855$ such signatures among our samples.
However, due to the measurement noise we set a more conservative filtering threshold of $4.76\times 10^8$ cycles, and obtained only $53$ high quality signatures. 
Experimentally, we observed that it took 34 signatures to recover the key with $100\%$ success rate. 
Running BKZ with block size 30 for the lattice of this size took 2 to 3 seconds on our experimental machine. 
After obtaining the key, we recovered the nonces and verified that most of them had the eight MSBs set to zero\footnote{There were few samples with 12 zero MSBs in the analysis}. 
If we had used the entire distribution we would need about $256\times34=8,704$ signatures.
We use the empirical numbers from our experiments to estimate the likelihood of obtaining such samples in our experimental setup given our choice of thresholds and the noise we experienced; in this case the probability of obtaining such a sample is 53/855.
The estimated total number of signatures required to carry out the attack is then 140,413, which takes about 163 minutes to collect. 
In the 4-bit case, the thresholds we used to filter the samples were those between $4.8\times 10^8$ and $4.81\times 10^8$ cycles.
With $77$ signatures we recover the key with overwhelming probability. This translates to $77\times 16=1,232$ signatures.
But we also need to account for filtering from a narrower range, which results in 1,121 samples out of the $13,687$ expected signatures with 4 LZBs from our total of 219,000 samples. 
In this case, we estimated that in total $15,042$ signatures are required for the attack, which takes approximately 18 minutes to collect. 
The key recovery success rate is shown in~\autoref{fig:user_results_udp}.

\subsection{Threat Model III: Remote Adversary} \label{sec:strongswan}
In this section, we demonstrate the viability of {\bf over the network attacks} from clients targeting a server assisted by an on-board TPM.
Specifically, we target StrongSwan, an open-source IPsec Virtual Private Network (VPN) software server.
To this end, we first profile a custom synthesized UDP client/server setup where we can minimize noise.
This allows to gauge processing and networking timings.
We later analyze the timing leakage as observed by a remote client from a server running StrongSwan VPN software.

\subsubsection{Remote UDP Attack}
\begin{figure}[t!]
    \centering
\includegraphics[width=0.90\linewidth]{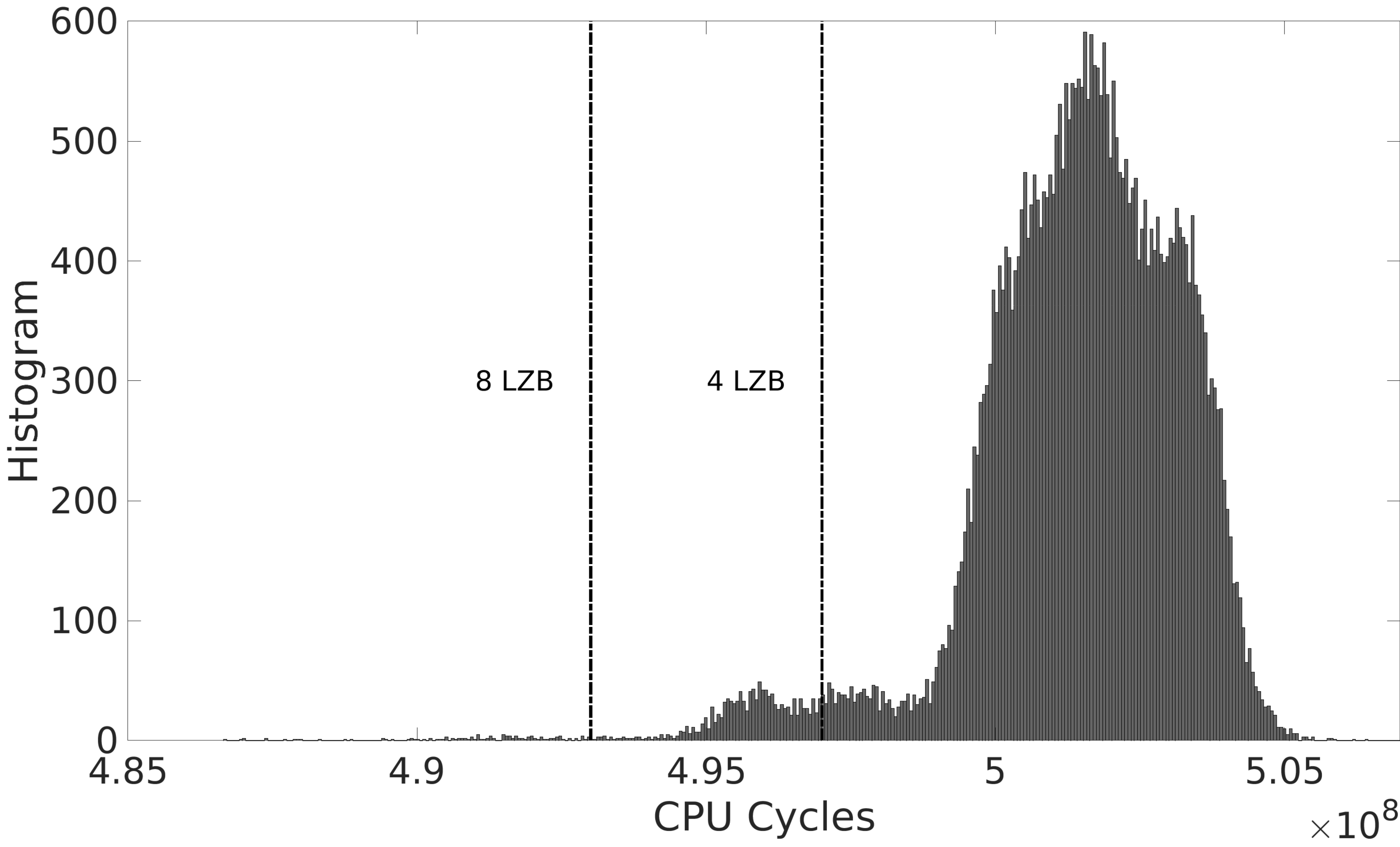} 
\caption{Histogram of ECDSA (NIST-256p) signature computation times over the network for 40,000 observations. 
A server application running on our Core i7-8705G machine is performing signing operations over a simple UDP-based protocol.
The client measures the request/response round-trip time to receive a new signature after each request.}
    \label{fig:histecdsaudp}
\end{figure}

We created a server application that uses the {\sf Intel \fTPM} to perform signing operations.
The server receives a request for a signature and returns the signature to the user over a simple protocol based on UDP.
The client (the attacker) sends requests to the server and collects the signatures, while timing the request/response round-trip time.  
~\autoref{fig:histecdsaudp} shows the collected timing information for 40,000 requests. 
Although there is some noise in the measurement, we can still distinguish signatures that are generated using short nonces. 
\autoref{fig:user_results_udp} shows our key recovery results.

The experimental results match our expectations outlined earlier, since the TPM takes around 200 milliseconds to generate a signature, 
which is a large enough window to leak timing information over the network. 
We filtered 8-bit samples by thresholding at $4.93\times 10^8$ cycles and for 4-bit samples at $4.97\times 10^8$ cycles measured on the client. 
For the case of 4-bit bias, we need 78 signatures above our timing threshold to recover the key, which corresponds to 1,248 signature operations by the server. 
This can be collected in less than 4 minutes. 
For the case of 8-bit bias we recover the key using 47 signatures with high probability, which requires 31 minutes of signing operations. 
These results demonstrate that remote attacks on \fTPM\ are viable. 
Next we explore this direction further by targeting the StrongSwan VPN product.

\subsubsection{Remote Timing Attack against StrongSwan}
StrongSwan is an open-source IPsec Virtual Private Network (VPN) implementation that is supported by modern OSes, including Linux and Microsoft Windows.
VPNs can use the IPSec protocol for encryption and authentication.
The IPsec key negotiation happens via the IKE protocol, which can use either pre-shared secrets or digital certificates for authentication.
StrongSwan further supports IKEv2 with signature-based authentication using a TPM 2.0 supported device~\cite{swTPM}.
Here, we attack a StrongSwan VPN Server that is configured to use the TPM for digital signature authentication by measuring the IKE authentication handshake. 

\begin{figure}[t!]
    \centering
\includegraphics[width=.9\linewidth]{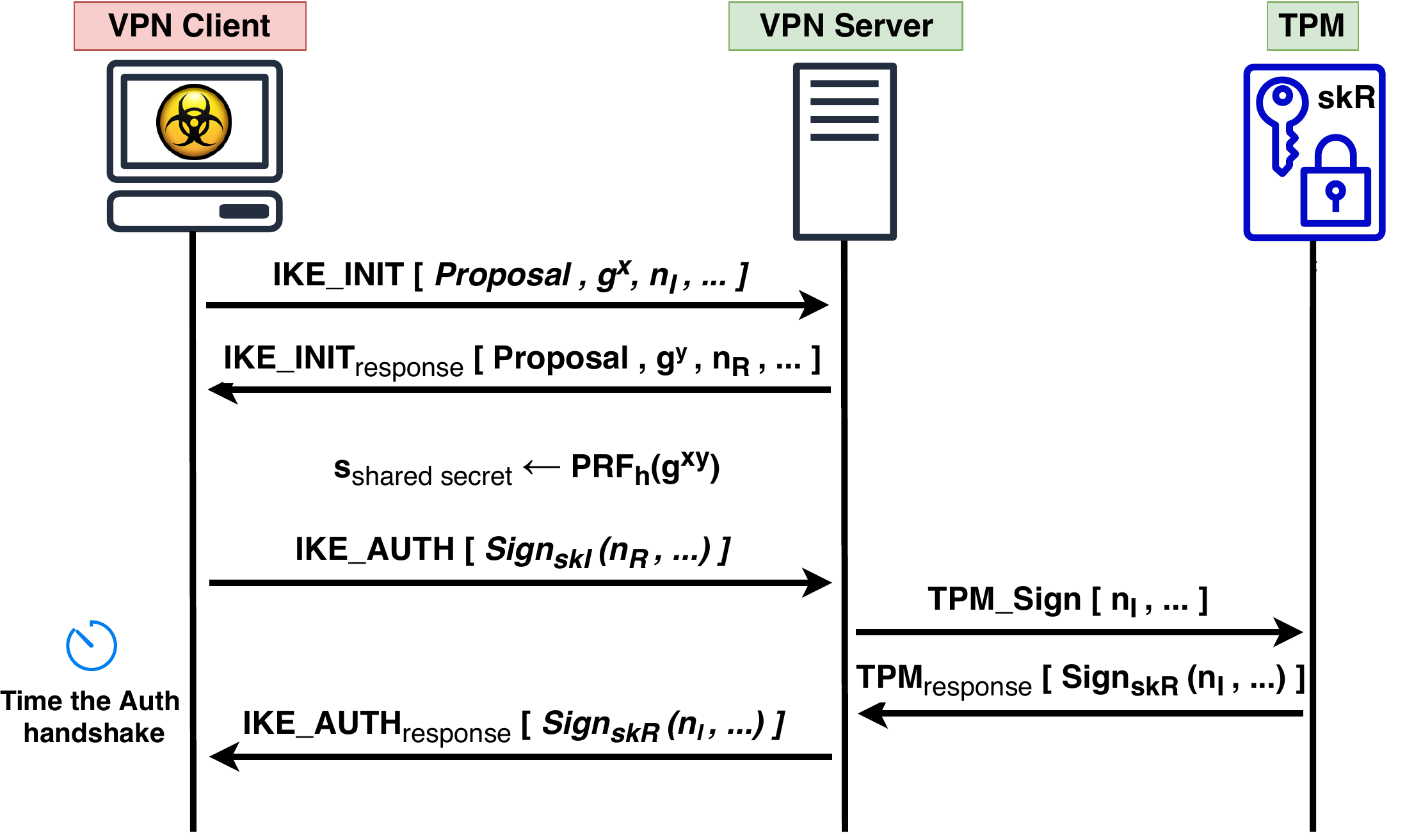} 
\caption{Steps of IKE\_SA\_INIT and IKE\_AUTH exchange between the client and server running StrongSwan VPN.}
    \label{fig:ikev2tpm}
\end{figure}

\noindent
{\bf IKEv2 Interleaved Authentication with TPM signatures:}
We configure our server to use the standard IKEv2 signature authentication with interleaved handshakes where the authentication is performed by an IKE\_SA\_INIT and a IKE\_AUTH exchange between the client and server.
\autoref{fig:ikev2tpm} shows these two handshakes, where the second handshake triggers the TPM device to sign the authentication message.
The first exchange of the IKE session, IKE\_SA\_INIT, negotiates security parameters, sends nonces and performs the Diffie-Hellman Key exchange.
After the first exchange, the second exchange, IKE\_AUTH, can be encrypted using the shared Diffie-Hellman (DH) key.
In the second exchange, the two parties verify each others' identities by signing each others' nonces.
We generated a unique ECDSA attestation key (AK) using the Intel \fTPM\ device on the VPN server.
The TPM device only exposes the public portion of the AK.
Then we generated a self-signed attestation identity key (AIK) certificate and stored the ECDSA AIK certificate in the non-volatile memory of the TPM device.
During the second exchange, the server asks the TPM device holding the private AK to sign the client's nonce and return the signature to the client.
When the client receives the signature, she can verify that her nonce is signed with the legitimate server's AK corresponding to the AIK certificate.
However, a malicious remote client, or a local user who can exploit the timing behavior to recover the private AK can forge valid signatures, and act as a legitimate VPN server.

\medskip
\noindent
{\bf StrongSwan VPN Key Recovery:}

\begin{figure}[t!]
    \centering
\includegraphics[width=0.90\linewidth]{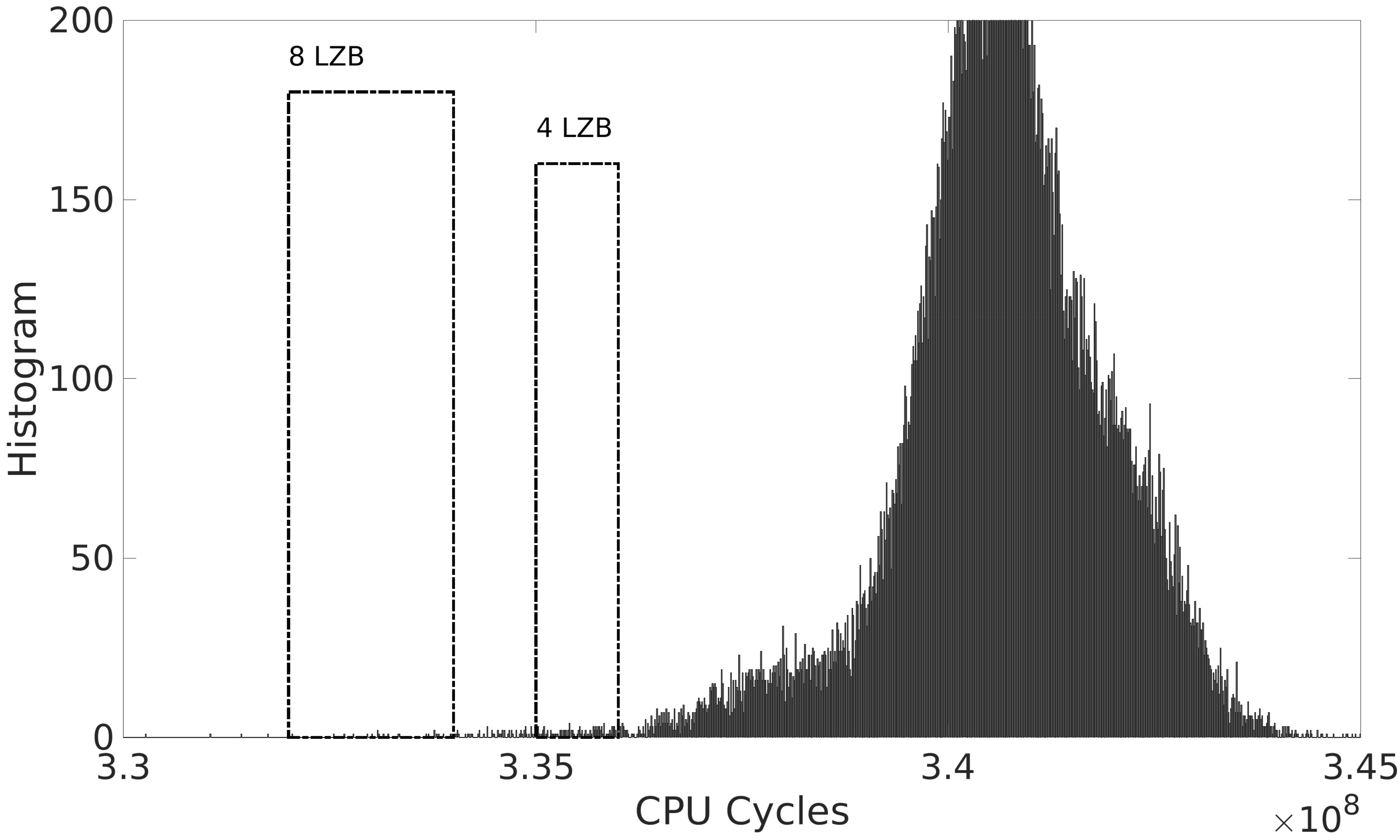}
\caption{Histogram of ECDSA (NIST-256p Curve) signing computation times over the network for 40K observations. The server is running StrongSwan VPN software equipped with Intel \fTPM. The client application measures the request/response round-trip time.}
    \label{fig:ecdsa_intel_strongswan}
\end{figure}

\begin{figure}[t!]
    \centering
\includegraphics[width=0.90\linewidth]{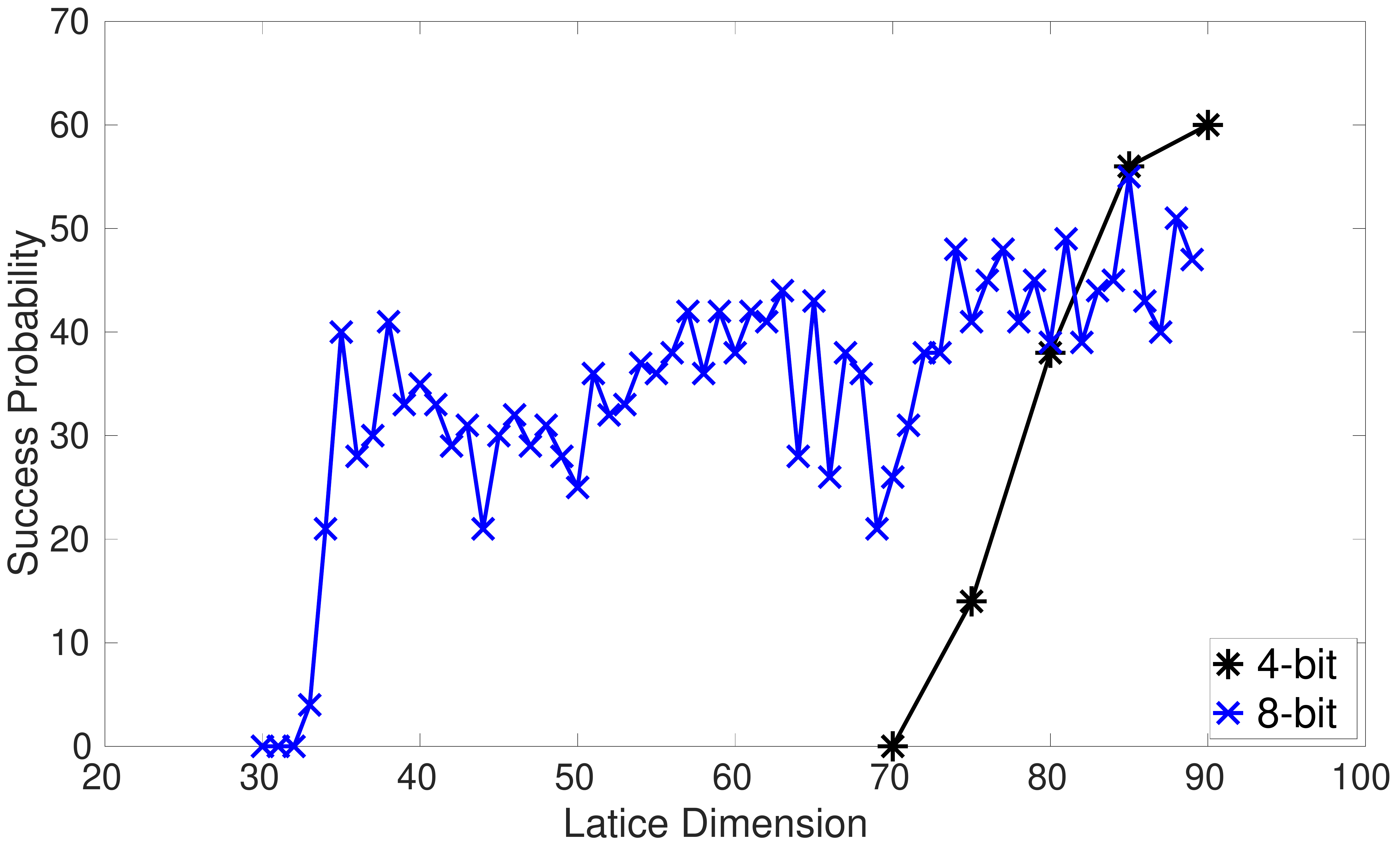}
\caption{{\bf Remote StrongSwan Attack:} Key recovery success probabilities by lattice dimension for the 4-bit and 8-bit cases for ECDSA (NIST-256p) with samples collected on the client.}
    \label{fig:swan_result}
\end{figure}

As a malicious client, we perform the following steps to collect timing measurement and recover the secret AK:
\begin{enumerate}
    \item The malicious client performs the first handshake with the server to exchange security parameters, nonces, and completes a Diffie-Hellman exchange.
    \item The malicious client starts a timer and initiates the second handshake. 
    After the server signs the client's nonce and other security parameters using the TPM device, the malicious client will receive the signature and measure the total handshake time. 
    The TPM signature timing vulnerability we discovered may delay this exchange based on the nonce used in signature generation, leaving an observable effect on network packet timings. 
    \item The malicious client stores the network timing and the received signature pairs and simply discards the session by sending an \texttt{IKE\_INFORMATION} packet to the server, 
    and it repeats this process starting from the first step to collect enough time measurements and signatures. 
\end{enumerate}
To determine if there is any exploitable leakage observed over the network, we collected both remote timings on the client and local timings on the server running a StrongSwan VPN software on our Core i7-8705G machine, where ECDSA signatures are computed by an Intel \fTPM.
The histograms for 40,000 timing measurements observed both locally and on the server are shown in~\autoref{fig:ecdsa_intel_local} and~\autoref{fig:ecdsa_intel_strongswan}. 
The clearly identifiable separate peaks corresponding to 4-bit and 8-bit leakage in~\autoref{fig:ecdsa_intel_local} are no longer observable with measurements collected over the noisy network in~\autoref{fig:ecdsa_intel_strongswan}.
Still, the relative location of the peaks in the local timings histogram can be used as a template to design filters to be applied on the remote timings. For this, we need to account for the change in clock frequencies. As a simple heuristic, we scale the filter ranges in~\autoref{fig:ecdsa_intel_local} by the ratio of the time when the largest peaks are observed, i.e. $3.41/4.82$. We also adjusted the filters to account for the additional delay due to remote measurements. Finally we reduced the widths to cover the left half of the distributions, since they yield cleaner samples.
For 8-bit samples we filter between $3.32\times 10^8$ and $3.34 \times 10^8$, and for 4-bit $3.35\times 10^8$ and $3.36\times 10^8$, obtaining 153 8-bit and 222 4-bit samples. We then applied the lattice attacks from~\autoref{sec:lattice} to these samples using our {\sf Sage} implementation and {\sf BKZ-2.0} reduction with block size 30 over many iterations.
The results are shown in the graph in~\autoref{fig:swan_result}.
For both the 4-bit and 8-bit cases, we recover the key with high probability after dimensions 34 and 80, respectively. 
In the 4-bit case we used 222 out of the expected $1/16\times 198K = 12,375$ 4-bit samples.
To end up with 80 4-bit samples we would need to samples $80\times 16=1,280$ samples.
However since we are filtering for high quality samples within the nonces with 4-bit bias with probability $222/12,375$ we need to also take that into account.
This means we need about $1,280\times 12,375/222 = 71,351$ signatures. 
In the 8-bit case used 153 out of the 774 expected 8-bit samples. 
This means we need about $34\times 256=8,704$ samples.
Accounting for filtering with probability $153/774$, we need about $8,704\times 774/153 = 44,032$ signatures.
In this case, targeting the nonces with 8-bit bias turns out to be more efficient, as the noise introduced by measuring remotely on the client side has rendered 4-bit samples harder to distinguish, and therefore these require more aggressive filtering.
We can collect about 139 signatures per minute from StrongSwan.
This means we can collect enough samples in about 5 hours 16 minutes.

In our attack, we queried the VPN server directly to collect the signatures and timings. 
This attack can also be performed by an active man-in-the-middle (MiTM) adversary who hijacks a DH key exchange. 
However, there is no additional benefit to be gained over the malicious client since the attacker is active in both scenarios.
A passive attack would not be possible, since the signatures are encrypted with the shared secret between the client and the server. 
Another important factor that affects the viability of the attack is networking noise. Depending on the type and traffic of the network, e.g. networks with high bandwidth, or local organizational networks and local private networks on the cloud, the success rate of the attack will vary. Typically in cloud environments, network connections between cloud nodes tend to have higher bandwidth and more stable connections, and thus will have less timing noise. 

\section{Discussion}
\medskip
\noindent
\textbf{Infineon ECDSA Timing Behavior:} 
\autoref{fig:ecdsa_inf} shows that the TPM manufactured by Infineon experiences non-constant-time behavior for ECDSA. 
We performed similar analysis by observing the correlation of LZBs in the nonce and timing (\autoref{fig:ecdsa_inf_boxplot}), and we did not observe any exploitable bias based on the timings.
We also performed other intuitive tests such as looking at the correlation between the timing behavior and the occurence of 1s. 
None of our tests were successful in finding time-dependent bias in the nonce.

\medskip
\noindent
\textbf{RSA Timing Behavior:} \label{sec:findings}
Using the methodology described in~\autoref{sec:timing}, we also profiled the timing behavior of the RSA signature scheme.
In~\autoref{tab:machines}, we report the average number of CPU cycles to compute RSA signatures for five configurations that support Intel \fTPM\ and three different configurations with a dedicated TPM chip.
For this test, we generated 40,000 valid 2048-bit RSA keys, programmed the TPM with these keys one at a time, and measured timings for RSA signing operations on the TPM.

The timing distributions for the dedicated TPM devices manufactured by Infineon and STMicroelectronics are fairly uniform, as shown in~\autoref{fig:rsa_stm} and~\autoref{fig:rsa_inf}.
In contrast, the distributions in~\autoref{fig:rsa_intel} and~\autoref{fig:rsa_nuv} show that RSA signature generation is not constant time on Intel \fTPM\ and the dedicated Nuvoton TPM; 
rather, it has a logarithmic timing distribution that depends on the key bits. 

This type of key-dependent timing behavior has previously been observed for the RSA implementation of Intel's IPP Cryptography library~\cite{wichelmann2018microwalk}. 
This implementation is based on the Chinese Remainder Theorem (CRT)~\cite{dingyi1996chinese}, and 
the timing variation is due to the modular inversion operation's use of the recursive Extended Euclidean Algorithm (EEA)\footnote{During disclosure Intel also confirmed that a version of the Intel IPP Cryptography library was running in Intel fTPM.}. 
After the CRT components of the signature are computed, the EEA is employed to compute the modular inverses that are needed to reconstruct the final signature.
EEA performs modular reductions using division and recurses according to the Euclidean algorithm until the remainder is zero. In this case, the observed timing behavior leaks the number of divisions.
Although we observe key-dependent leakage, the EEA algorithm operates serially, and we may only recover a few initial bits of independent RSA keys.  This does not seem to leak enough information to recover the full RSA keys using lattice-based or similar methods, which require a larger proportion of known bits of the secret key for full RSA key recovery.~\cite{coppersmith1997small}
\begin{table}[t!]
\centering
 	\caption{Summary of our key recovery results.}
     \label{tab:results}
\begin{tabular}{llccc} 
\toprule
\footnotesize{\textbf{Threat Model}}              & \footnotesize{\textbf{TPM}}        & \footnotesize{\textbf{Scheme}} & \footnotesize{\textbf{\#Sign.}} & \footnotesize{\textbf{Time}} \\ 
\midrule 
\footnotesize{Local System}      &  \footnotesize{ST TPM} & \footnotesize{ECDSA}  & \footnotesize{39,980} & \footnotesize{80 mins} \\
\footnotesize{Local System}      &  \footnotesize{\fTPM} & \footnotesize{ECDSA}  & \footnotesize{1,248} & \footnotesize{4 mins} \\
\footnotesize{Local System}      &  \footnotesize{\fTPM} & \footnotesize{ECSchnorr} & \footnotesize{1,040} & \footnotesize{3 mins} \\
\footnotesize{Local User  }      &  \footnotesize{\fTPM} & \footnotesize{ECDSA}  & \footnotesize{15,042} & \footnotesize{18 mins} \\ 
\footnotesize{Remote SSwan}      &  \footnotesize{\fTPM} & \footnotesize{ECDSA}  & \footnotesize{44,032} & \footnotesize{$\sim$5 hrs} \\ 
\bottomrule
\end{tabular}
\end{table}

\section{Countermeasures}
Software-based countermeasures can be temporarily deployed to mitigate the user and network attacks we discuss.
The OS can add a pre-determined delay to the TPM interface for TPM commands to ensure that it is executed in constant time.
However, this requires precise estimation of an upper bound for the execution time for these operations.
This is not trivial, since the execution times vary among different TPMs.
An intrusion detection (IDS) system may also be able to detect such attacks by inspecting API calls and/or network traffic. 
However, IDS rules can be avoided in many cases by determined adversaries, and they may suffer from false positives.
For example, an adversary can introduce random delay between requests or combine the malicious requests with benign ones to circumvent detection.

Constant-time implementation techniques are known, but these incur additional development and execution costs.
The standard defense is to deploy these techniques as firmware and software patches, or to replace the vulnerable TPM when patching is not feasible. 
Intel has promised patches for Intel fTPM, which is executed as part of the Intel Management Engine.
We have also shared our tools and techniques with ST, and they are evaluating new versions of their products based on our findings. 
It is important that these countermeasures do not compromise the randomness and uniformity of the ECDSA nonce~\cite{ubikeyvuln}.

\section{Conclusions}
Since TPMs act as a root of trust, most physical TPMs have undergone validation through FIPS 140-2, 
which includes physical protection, as well as the more rigorous certification based on Common Criteria up to levels of EAL 4+. 
This certification is intended to ensure protection against a wide range of attacks, including physical and side-channel attacks against its cryptographic capabilities. 
However, this is the second time that the CC evaluation process has failed to provide expected security guarantees~\cite{nemec2017return}.
This clearly underscores the need to reevaluate the CC process. 
Given the rapid proliferation of side-channel attacks, it would be advisable to switch to a continuously evolving evaluation process. 
We also note that another potentially vulnerable trusted platform is a Hardware Security Module (HSM). 
Recent works have already demonstrated that HSMs have more severe vulnerabilities \cite{BG19}. 
We expect HSMs to have similar security issues, since most have not even been certified or tested by an external authority. 

The vulnerabilities discovered in this paper apply to a wide range of computing devices.
The vulnerable {\bf Intel \fTPM} is used by many PC and laptop manufacturers, including Lenovo, Dell, and HP.
Many new laptop manufacturers prefer using the integrated Intel \fTPM\ rather than adding extra hardware.
The Intel \fTPM\ is somewhat comparable to a hardware TPM since it isolates execution in an isolated 32-bit microcontroller. It is also widely used by the Intel IoT platform.
Our results on the ST TPM, however, show that even OEMs making a conservative choice and trusting CC-certified hardware TPMs may fall victim to side-channel key recovery attacks. 
More specifically, we demonstrated vulnerabilities in Intel \fTPM\ and STMicroelectronics TPM devices.
We found additional non-constant execution timing leakage in Infineon and Nuvoton TPMs.
Concretely, we managed to recover ECDSA and ECSchnorr keys by collecting signature timing data with and without administrative privileges. Further, we managed to recover ECDSA keys from a \fTPM-endowed server running StrongSwan VPN over a noisy network as measured by a client. The fact that a remote attack can extract keys from a TPM device certified as secure against side-channel leakage underscores the need to reassess remote attacks on cryptographic implementations, which had been considered a solved problem.

\subsection*{Acknowledgments}
We thank Lejla Batina and the anonymous reviewers for their valuable comments for improving the quality of this paper.

This work was supported by the National Science Foundation under grants no. CNS-1513671, CNS-1651344, and CNS-1814406.
Additional funding was provided by a generous gift from Intel.
Heninger performed some of this research while visiting Microsoft Research New England.

{\footnotesize \bibliographystyle{plain}
\bibliography{ref}

\begin{thebibliography}{10}

\bibitem{al2013lucky}
N.~J. {Al Fardan} and K.~G. {Paterson}.
\newblock {Lucky Thirteen: Breaking the TLS and DTLS Record Protocols}.
\newblock In {\em 2013 IEEE Symposium on Security and Privacy}, pages 526--540,
  May 2013.

\bibitem{arm2009security}
A~ARM.
\newblock {Security technology building a secure system using trustzone
  technology (white paper)}.
\newblock {\em ARM Limited}, 2009.

\bibitem{bajikar2002trusted}
Sundeep Bajikar.
\newblock {Trusted Platform Module (TPM) based security on notebook pcs-white
  paper}.
\newblock {\em Mobile Platforms Group Intel Corporation}, 1:20, 2002.

\bibitem{barreto2005pairing}
Paulo S. L.~M. Barreto and Michael Naehrig.
\newblock {Pairing-Friendly Elliptic Curves of Prime Order}.
\newblock In Bart Preneel and Stafford Tavares, editors, {\em Selected Areas in
  Cryptography}, pages 319--331, Berlin, Heidelberg, 2006. Springer Berlin
  Heidelberg.

\bibitem{benger2014ooh}
Naomi Benger, Joop van~de Pol, Nigel~P. Smart, and Yuval Yarom.
\newblock {``Ooh Aah... Just a Little Bit'' : A Small Amount of Side Channel
  Can Go a Long Way}.
\newblock In Lejla Batina and Matthew Robshaw, editors, {\em Cryptographic
  Hardware and Embedded Systems -- CHES 2014}, pages 75--92, Berlin,
  Heidelberg, 2014. Springer Berlin Heidelberg.

\bibitem{bleichenbacher2005experiments}
Daniel Bleichenbacher.
\newblock Experiments with dsa.
\newblock {\em CRYPTO 2005--Rump Session}, 2005.

\bibitem{boneh2001modular}
Dan Boneh, Shai Halevi, and Nick Howgrave-Graham.
\newblock {The modular inversion hidden number problem}.
\newblock In {\em International Conference on the Theory and Application of
  Cryptology and Information Security}, pages 36--51. Springer, 2001.

\bibitem{boneh1996hardness}
Dan Boneh and Ramarathnam Venkatesan.
\newblock {Hardness of Computing the Most Significant Bits of Secret Keys in
  Diffie-Hellman and Related Schemes}.
\newblock In Neal Koblitz, editor, {\em Advances in Cryptology --- CRYPTO '96},
  pages 129--142, Berlin, Heidelberg, 1996. Springer Berlin Heidelberg.

\bibitem{brumley2011remote}
Billy~Bob Brumley and Nicola Tuveri.
\newblock {Remote Timing Attacks Are Still Practical}.
\newblock In Vijay Atluri and Claudia Diaz, editors, {\em Computer Security --
  ESORICS 2011}, pages 355--371, Berlin, Heidelberg, 2011. Springer Berlin
  Heidelberg.

\bibitem{brumley2005remote}
David Brumley and Dan Boneh.
\newblock {Remote timing attacks are practical}.
\newblock {\em Computer Networks}, 48(5):701 -- 716, 2005.
\newblock Web Security.

\bibitem{butterworth2013bios}
John Butterworth, Corey Kallenberg, Xeno Kovah, and Amy Herzog.
\newblock Bios chronomancy: Fixing the core root of trust for measurement.
\newblock In {\em Proceedings of the 2013 ACM SIGSAC Conference on Computer
  \&\#38; Communications Security}, CCS '13, pages 25--36, New York, NY, USA,
  2013. ACM.

\bibitem{Challener2011}
David Challener.
\newblock {\em {{Trusted Platform Module}}}, pages 1332--1335.
\newblock Springer US, Boston, MA, 2011.

\bibitem{coppersmith1997small}
Don Coppersmith.
\newblock Small solutions to polynomial equations, and low exponent rsa
  vulnerabilities.
\newblock {\em Journal of Cryptology}, 10(4):233--260, 1997.

\bibitem{dall2018cachequote}
Fergus Dall, Gabrielle De~Micheli, Thomas Eisenbarth, Daniel Genkin, Nadia
  Heninger, Ahmad Moghimi, and Yuval Yarom.
\newblock {Cachequote: Efficiently recovering long-term secrets of sgx epid via
  cache attacks}.
\newblock {\em IACR Transactions on Cryptographic Hardware and Embedded
  Systems}, pages 171--191, 2018.

\bibitem{de2014using}
Elke De~Mulder, Michael Hutter, Mark~E. Marson, and Peter Pearson.
\newblock {Using Bleichenbacher's solution to the hidden number problem to
  attack nonce leaks in 384-bit ECDSA: extended version}.
\newblock {\em Journal of Cryptographic Engineering}, 4(1):33--45, Apr 2014.

\bibitem{den2002dpa}
Bert den Boer, Kerstin Lemke, and Guntram Wicke.
\newblock {A DPA attack against the modular reduction within a CRT
  implementation of RSA}.
\newblock In {\em International Workshop on Cryptographic Hardware and Embedded
  Systems}, pages 228--243. Springer, 2002.

\bibitem{dingyi1996chinese}
Pei Dingyi, Salomaa Arto, and Ding Cunsheng.
\newblock {\em Chinese remainder theorem: applications in computing, coding,
  cryptography}.
\newblock World Scientific, 1996.

\bibitem{ermolov2017hack}
Mark Ermolov and Maxim Goryachy.
\newblock {How to hack a turned-off computer, or running unsigned code in intel
  management engine}.
\newblock {\em Black Hat Europe}, 2017.

\bibitem{Ermolov2017JTAG}
Mark Ermolov and Maxim Goryachy.
\newblock {Where There's a JTAG, There's a way: Obtaining full system access
  via USB}.
\newblock {\em White Paper}, 2017.
\newblock Accessed: \today.

\bibitem{ermolov2019hack}
Mark Ermolov and Maxim Goryachy.
\newblock {Intel VISA: Through the Rabbit Hole}.
\newblock {\em Black Hat Asia}, 2019.

\bibitem{fan2016attacking}
Shuqin Fan, Wenbo Wang, and Qingfeng Cheng.
\newblock {Attacking OpenSSL Implementation of ECDSA with a Few Signatures}.
\newblock In {\em Proceedings of the 2016 ACM SIGSAC Conference on Computer and
  Communications Security}, CCS '16, pages 1505--1515, New York, NY, USA, 2016.
  ACM.

\bibitem{gallagher2013digital}
Patrick Gallagher.
\newblock {Digital signature standard (DSS)}.
\newblock {\em Federal Information Processing Standards Publications, volume
  FIPS}, pages 186--3, 2013.

\bibitem{googlevtpm}
{Google}.
\newblock {Shielded VM}.
\newblock
  \url{https://cloud.google.com/security/shielded-cloud/shielded-vm#vtpm},
  2019.
\newblock Accessed: \today.

\bibitem{han2018bad}
Seunghun Han, Wook Shin, Jun-Hyeok Park, and HyoungChun Kim.
\newblock {A Bad Dream: Subverting Trusted Platform Module While You Are
  Sleeping}.
\newblock In {\em 27th {USENIX} Security Symposium ({USENIX} Security 18)},
  pages 1229--1246, Baltimore, MD, August 2018. {USENIX} Association.

\bibitem{hlavavc2006extended}
Martin Hlav{\'a}{\v{c}} and Tom{\'a}{\v{s}} Rosa.
\newblock {Extended Hidden Number Problem and Its Cryptanalytic Applications}.
\newblock In Eli Biham and Amr~M. Youssef, editors, {\em Selected Areas in
  Cryptography}, pages 114--133, Berlin, Heidelberg, 2007. Springer Berlin
  Heidelberg.

\bibitem{howgrave2001lattice}
Nick~A Howgrave-Graham and Nigel~P. Smart.
\newblock {Lattice attacks on digital signature schemes}.
\newblock {\em Designs, Codes and Cryptography}, 23(3):283--290, 2001.

\bibitem{intelIPP}
{Intel}.
\newblock {Developer Reference for Intel Integrated Performance Primitives
  Cryptography}.
\newblock \url{https://software.intel.com/en-us/ipp-crypto-reference}, 2019.
\newblock Accessed: \today.

\bibitem{intelQuark}
{Intel}.
\newblock {Intel Quark Microcontrollers}.
\newblock
  \url{https://www.intel.com/content/www/us/en/embedded/products/quark/overview.html},
  2019.
\newblock Accessed: \today.

\bibitem{BG19}
Gabriel~Campana Jean-Baptiste~Bedrune.
\newblock {Everybody be Cool, This is a Robbery!}
\newblock {\em Black Hat USA}, 2019.

\bibitem{johnson2001elliptic}
Don Johnson, Alfred Menezes, and Scott Vanstone.
\newblock {The Elliptic Curve Digital Signature Algorithm (ECDSA)}.
\newblock {\em International Journal of Information Security}, 1(1):36--63, Aug
  2001.

\bibitem{kauer2007oslo}
Bernhard Kauer.
\newblock {OSLO: Improving the Security of Trusted Computing.}
\newblock In {\em USENIX Security Symposium}, pages 229--237, 2007.

\bibitem{kocher1996timing}
Paul~C. Kocher.
\newblock {Timing Attacks on Implementations of Diffie-Hellman, RSA, DSS, and
  Other Systems}.
\newblock In Neal Koblitz, editor, {\em Advances in Cryptology --- CRYPTO '96},
  pages 104--113, Berlin, Heidelberg, 1996. Springer Berlin Heidelberg.

\bibitem{kursawe2005analyzing}
Klaus Kursawe, Dries Schellekens, and Bart Preneel.
\newblock {Analyzing trusted platform communication}.
\newblock In {\em In: ECRYPT Workshop, CRASH – CRyptographic Advances in
  Secure Hardware}, page~8, 2005.

\bibitem{LLL82}
A.~K. Lenstra, H.~W. Lenstra, and L.~Lovasz.
\newblock Factoring polynomials with rational coefficients.
\newblock {\em MATH. ANN}, 261:515--534, 1982.

\bibitem{mangard2008power}
Stefan Mangard, Elisabeth Oswald, and Thomas Popp.
\newblock {\em {Power analysis attacks: Revealing the secrets of smart cards}},
  volume~31.
\newblock Springer Science \& Business Media, 2008.

\bibitem{messerges2002examining}
Thomas~S. Messerges, Ezzat~A. Dabbish, and Robert~H. Sloan.
\newblock {Examining Smart-Card Security Under the Threat of Power Analysis
  Attacks}.
\newblock {\em IEEE Trans. Comput.}, 51(5):541--552, May 2002.

\bibitem{meyer2014revisiting}
Christopher Meyer, Juraj Somorovsky, Eugen Weiss, J{\"o}rg Schwenk, Sebastian
  Schinzel, and Erik Tews.
\newblock {Revisiting SSL/TLS Implementations: New Bleichenbacher Side Channels
  and Attacks}.
\newblock In {\em 23rd {USENIX} Security Symposium ({USENIX} Security 14)},
  pages 733--748, San Diego, CA, August 2014. {USENIX} Association.

\bibitem{howWindowsTPM}
{Microsoft}.
\newblock {How Windows 10 uses the Trusted Platform Module}.
\newblock
  \url{https://docs.microsoft.com/en-us/windows/security/information-protection/tpm/how-windows-uses-the-tpm},
  2019.
\newblock Accessed: \today.

\bibitem{azurevtpm}
{Microsoft}.
\newblock {Support for generation 2 VMs (preview) on Azure}.
\newblock
  \url{https://docs.microsoft.com/en-us/azure/virtual-machines/windows/generation-2},
  2019.
\newblock Accessed: \today.

\bibitem{mitchell2005trusted}
Chris Mitchell.
\newblock {\em {Trusted computing}}, volume~6.
\newblock Iet, 2005.

\bibitem{nemec2017return}
Matus Nemec, Marek Sys, Petr Svenda, Dusan Klinec, and Vashek Matyas.
\newblock {The Return of Coppersmith's Attack: Practical Factorization of
  Widely Used RSA Moduli}.
\newblock In {\em Proceedings of the 2017 ACM SIGSAC Conference on Computer and
  Communications Security}, CCS '17, pages 1631--1648, New York, NY, USA, 2017.
  ACM.

\bibitem{nguyen2002insecurity}
Nguyen and Shparlinski.
\newblock {The Insecurity of the Digital Signature Algorithm with Partially
  Known Nonces}.
\newblock {\em Journal of Cryptology}, 15(3):151--176, Jun 2002.

\bibitem{nguyen2003insecurity}
Phong~Q. Nguyen and Igor~E. Shparlinski.
\newblock {The Insecurity of the Elliptic Curve Digital Signature Algorithm
  with Partially Known Nonces}.
\newblock {\em Designs, Codes and Cryptography}, 30(2):201--217, Sep 2003.

\bibitem{lllontheaverage}
Phong~Q. Nguyen and Damien Stehl{\'e}.
\newblock {LLL} on the average.
\newblock In Florian Hess, Sebastian Pauli, and Michael Pohst, editors, {\em
  Algorithmic Number Theory}, pages 238--256, Berlin, Heidelberg, 2006.
  Springer Berlin Heidelberg.

\bibitem{pereida2016make}
Cesar Pereida~Garc\'{\i}a, Billy~Bob Brumley, and Yuval Yarom.
\newblock {Make Sure DSA Signing Exponentiations Really Are Constant-Time}.
\newblock In {\em Proceedings of the 2016 ACM SIGSAC Conference on Computer and
  Communications Security}, CCS '16, pages 1639--1650, New York, NY, USA, 2016.
  ACM.

\bibitem{perez2006vtpm}
Ronald Perez, Reiner Sailer, Leendert van Doorn, et~al.
\newblock {vTPM: virtualizing the trusted platform module}.
\newblock In {\em Proc. 15th Conf. on USENIX Security Symposium}, pages
  305--320, 2006.

\bibitem{quisquater2001electromagnetic}
Jean-Jacques Quisquater and David Samyde.
\newblock {ElectroMagnetic Analysis (EMA): Measures and Counter-measures for
  Smart Cards}.
\newblock In Isabelle Attali and Thomas Jensen, editors, {\em Smart Card
  Programming and Security}, pages 200--210, Berlin, Heidelberg, 2001. Springer
  Berlin Heidelberg.

\bibitem{raj2016ftpm}
Himanshu Raj, Stefan Saroiu, Alec Wolman, Ronald Aigner, Jeremiah Cox, Paul
  England, Chris Fenner, Kinshuman Kinshumann, Jork Loeser, Dennis Mattoon,
  Magnus Nystrom, David Robinson, Rob Spiger, Stefan Thom, and David Wooten.
\newblock {fTPM: A Software-Only Implementation of a {TPM} Chip}.
\newblock In {\em 25th {USENIX} Security Symposium ({USENIX} Security 16)},
  pages 841--856, Austin, TX, August 2016. {USENIX} Association.

\bibitem{romer2001information}
Tanja R{\"o}mer and Jean-Pierre Seifert.
\newblock Information leakage attacks against smart card implementations of the
  elliptic curve digital signature algorithm.
\newblock In {\em International Conference on Research in Smart Cards}, pages
  211--219. Springer, 2001.

\bibitem{ronen20199}
Eyal Ronen, Robert Gillham, Daniel Genkin, Adi Shamir, David Wong, and Yuval
  Yarom.
\newblock {The 9 Lives of Bleichenbacher’s CAT: New Cache ATtacks on TLS
  Implementations}.
\newblock In {\em IEEE Symposium on Security and Privacy}, 2019.

\bibitem{ryan2019return}
Keegan Ryan.
\newblock {Return of the Hidden Number Problem.}
\newblock {\em IACR Transactions on Cryptographic Hardware and Embedded
  Systems}, pages 146--168, 2019.

\bibitem{BKZ87}
C.~P. Schnorr.
\newblock A hierarchy of polynomial time lattice basis reduction algorithms.
\newblock {\em Theor. Comput. Sci.}, 53(2-3):201--224, August 1987.

\bibitem{schnorr1991efficient}
C.~P. Schnorr.
\newblock {Efficient signature generation by smart cards}.
\newblock {\em Journal of Cryptology}, 4(3):161--174, Jan 1991.

\bibitem{skochinsky2014intel}
Igor Skochinsky.
\newblock {Intel ME Secrets}.
\newblock {\em Code Blue}, 2014.

\bibitem{sparks2007security}
Evan~R Sparks and Evan~R Sparks.
\newblock {A security assessment of trusted platform modules computer science
  technical report TR2007-597}.
\newblock {\em Dept. Comput. Sci., Dartmouth College, Hanover, NH, USA, Tech.
  Rep., TR2007-597}, 2007.

\bibitem{STtpmCCcert}
{ST Microelectronics}.
\newblock {CC for IT security evaluation: Trusted Platform Module ST33TPHF2E
  mode TPM2.0}.
\newblock
  \url{https://www.ssi.gouv.fr/uploads/2018/10/anssi-cible-cc-2018_41en.pdf},
  2019.
\newblock Accessed: \today.

\bibitem{STtpmBrief}
{ST Microelectronics}.
\newblock {ST33TPHF2ESPI Product Brief}.
\newblock \url{https://www.st.com/resource/en/data_brief/st33tphf2espi.pdf},
  2019.
\newblock Accessed: \today.

\bibitem{swTPM}
{strongSwan}.
\newblock {Trusted Platform Module 2.0 - strongSwan}.
\newblock \url{https://wiki.strongswan.org/projects/strongswan/wiki/TpmPlugin},
  2019.
\newblock Accessed: \today.

\bibitem{tcg2005client}
PC~TCG.
\newblock {Client Specific-TPM Interface Specification (TIS) Version 1.2}.
\newblock {\em Trusted Computing Group}, 2005.

\bibitem{tcgCRB}
PC~TCG.
\newblock {TPM 2.0 Mobile Command Response Buffer Interface}.
\newblock {\em Trusted Computing Group}, 2014.

\bibitem{sagemath}
{The Sage Developers}.
\newblock {\em {S}ageMath, the {S}age {M}athematics {S}oftware {S}ystem
  ({V}ersion 8.4)}, 2019.
\newblock {\tt https://www.sagemath.org}.

\bibitem{TCGPP}
{Trusted Computing Group}.
\newblock {Protection Profile PC Client Specific TPM}.
\newblock
  \url{https://trustedcomputinggroup.org/wp-content/uploads/TCG_PP_PCClient_Specific_TPM2.0_v1.1_r1.38.pdf},
  2019.
\newblock Accessed: \today.

\bibitem{tpm2LibSpec}
{Trusted Computing Group}.
\newblock {TPM 2.0 Library Specification}.
\newblock
  \url{https://trustedcomputinggroup.org/resource/tpm-library-specification/},
  2019.
\newblock Accessed: \today.

\bibitem{ververis2010security}
Vassilios Ververis.
\newblock {Security evaluation of Intel's active management technology}, 2010.

\bibitem{wichelmann2018microwalk}
Jan Wichelmann, Ahmad Moghimi, Thomas Eisenbarth, and Berk Sunar.
\newblock Microwalk: A framework for finding side channels in binaries.
\newblock In {\em Proceedings of the 34th Annual Computer Security Applications
  Conference}, pages 161--173. ACM, 2018.

\bibitem{wong2015timing}
David Wong.
\newblock {Timing and Lattice Attacks on a Remote ECDSA OpenSSL Server: How
  Practical Are They Really?}
\newblock {\em IACR Cryptology ePrint Archive}, 2015:839, 2015.

\bibitem{ubikeyvuln}
{Yubico}.
\newblock {Security Advisory 2019-06-13 – Reduced initial randomness on FIPS
  keys}.
\newblock
  \url{https://www.yubico.com/support/security-advisories/ysa-2019-02/}, 2019.
\newblock Accessed: \today.

\end{thebibliography}

\appendix

\section{Additional Timing Analysis Figures}
\begin{figure}[htp]
    \centering
\includegraphics[width=0.90\linewidth]{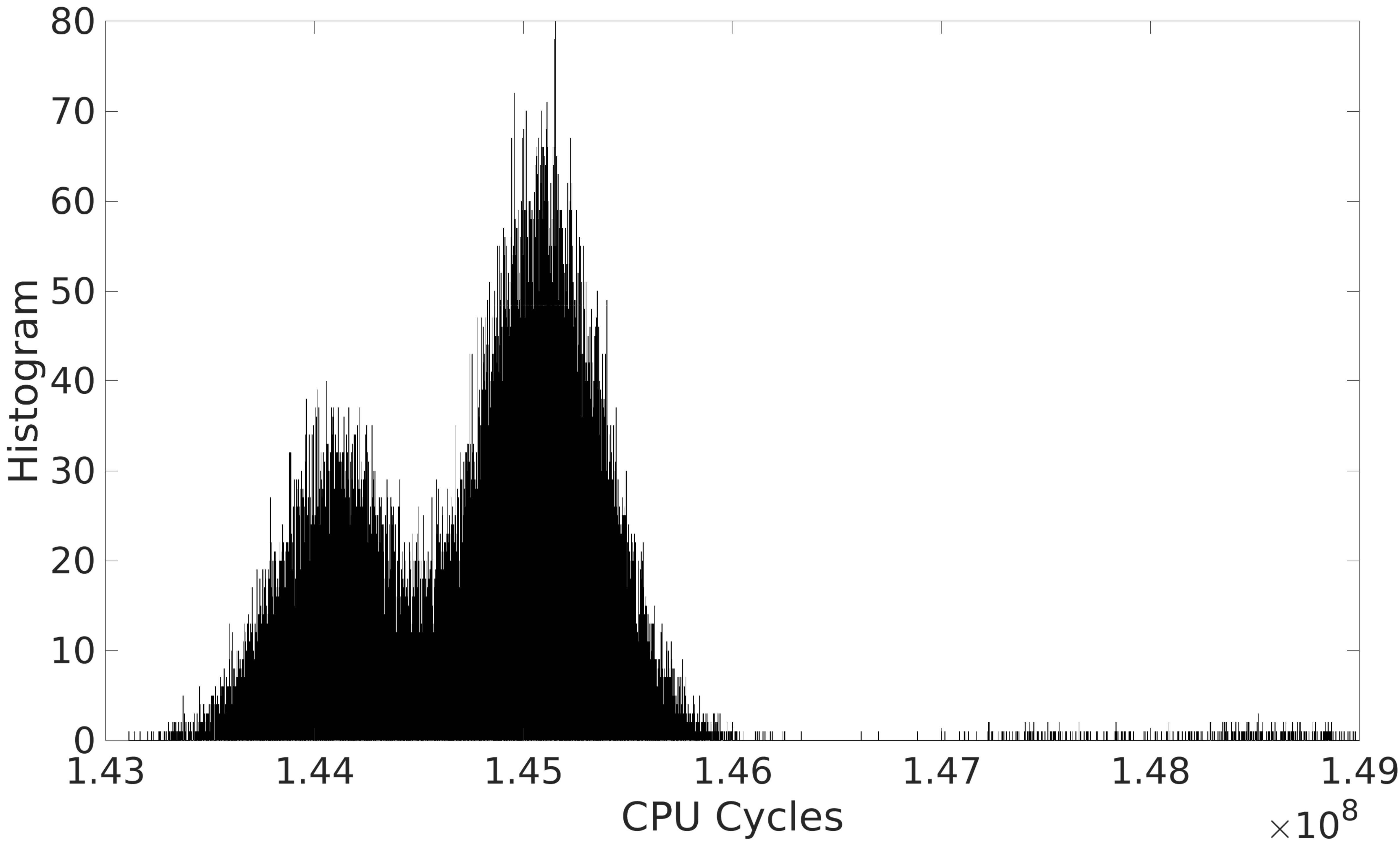} 
\caption{Histogram of ECDSA (NIST-256p) signature generation timings a dedicated Infineon TPM as measured on a Core i7-8650U machine for 40,000 observations.}
    \label{fig:ecdsa_inf}
\end{figure}

\begin{figure}[htp]
    \centering
\includegraphics[width=0.90\linewidth]{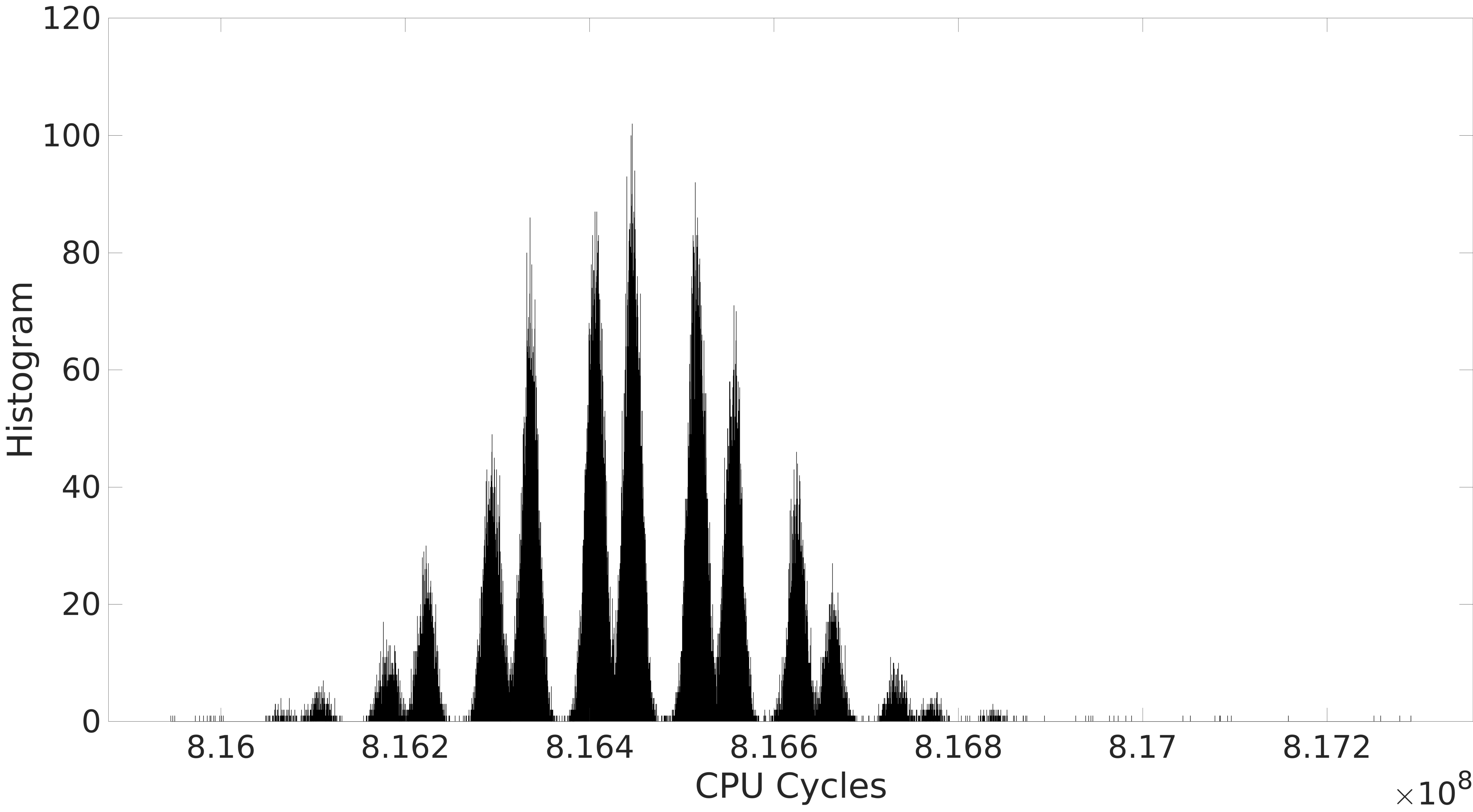} 
\caption{Histogram of RSA-2048 signature generation timings on Intel \fTPM\ as measured on a Core i7-7700 machine for 40,000 observations.}
    \label{fig:rsa_intel}
\end{figure}

\begin{figure}[htp]
    \centering
\includegraphics[width=0.90\linewidth]{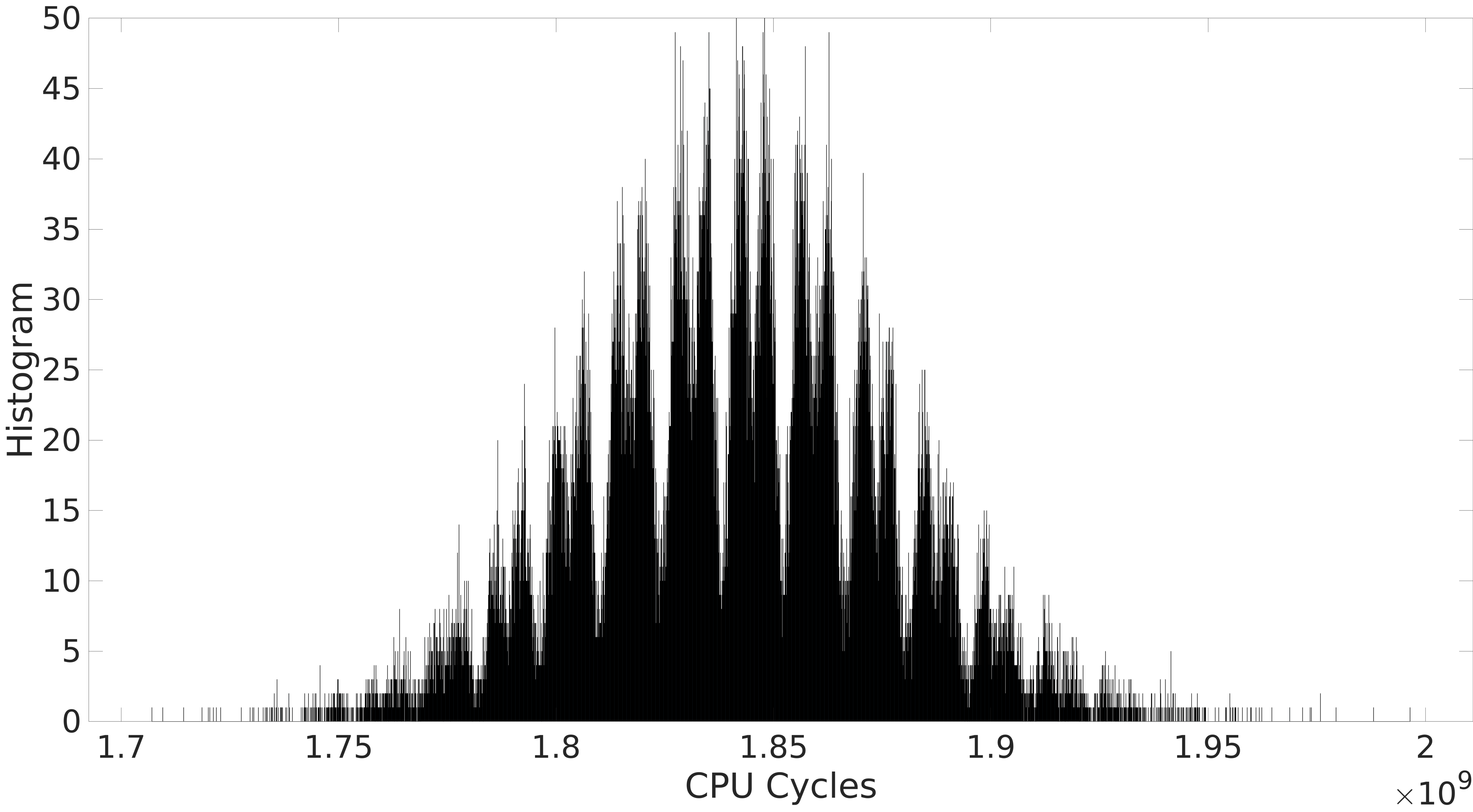}
\caption{Histogram of RSA-2048 signature generation timings on a dedicated Nuvoton TPM as measured on a Core i5-6440HQ machine for 40,000 observations.}
    \label{fig:rsa_nuv}
\end{figure}

 \begin{figure}[htp]
     \centering
 \includegraphics[width=0.90\linewidth]{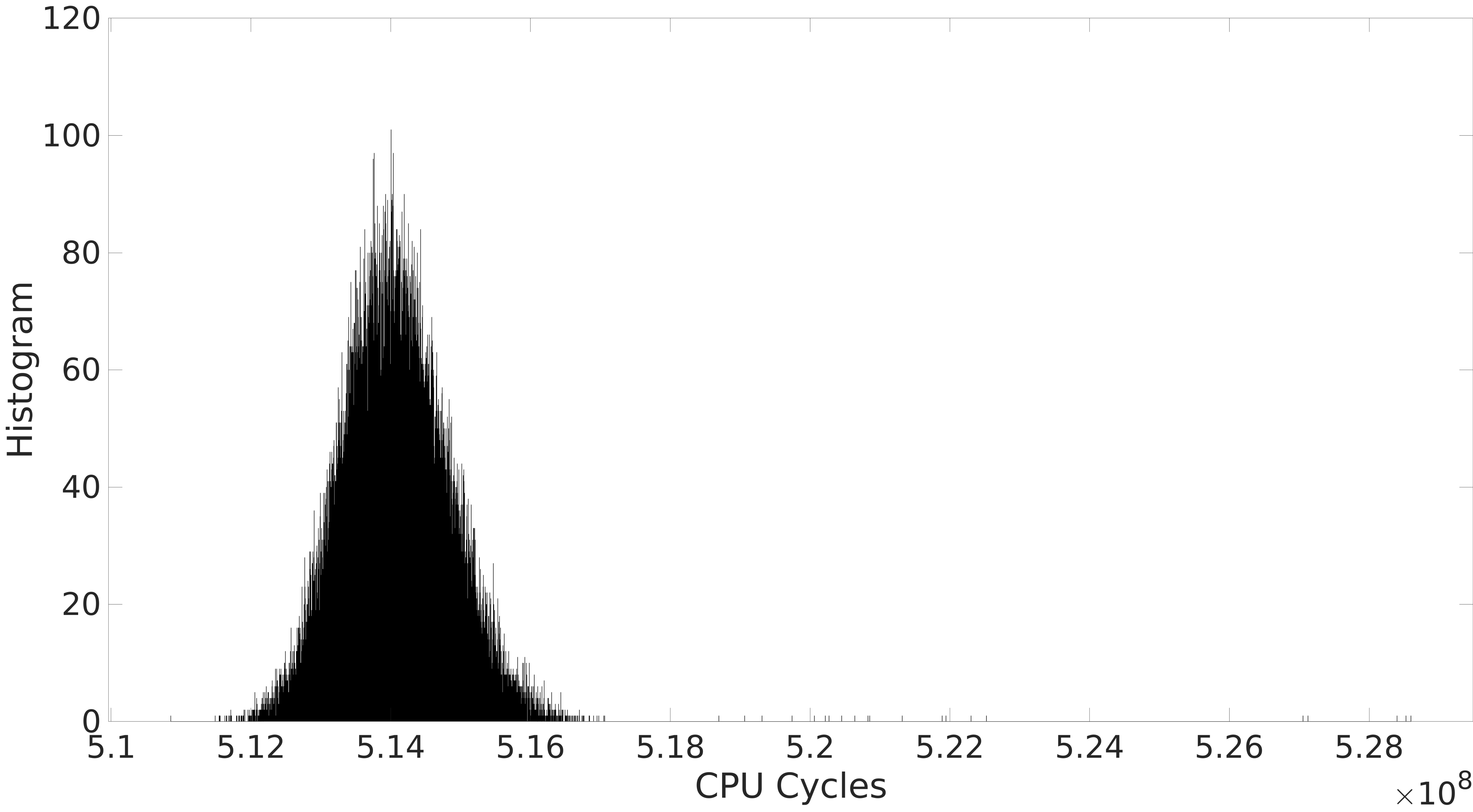}
 \caption{Histogram of RSA-2048 signature generation timings on a dedicated STMicroelectronics TPM as measured on a Core i7-8650U machine for 40,000 observations.}
     \label{fig:rsa_stm}
 \end{figure}

 \begin{figure}[htp]
     \centering
 \includegraphics[width=0.90\linewidth]{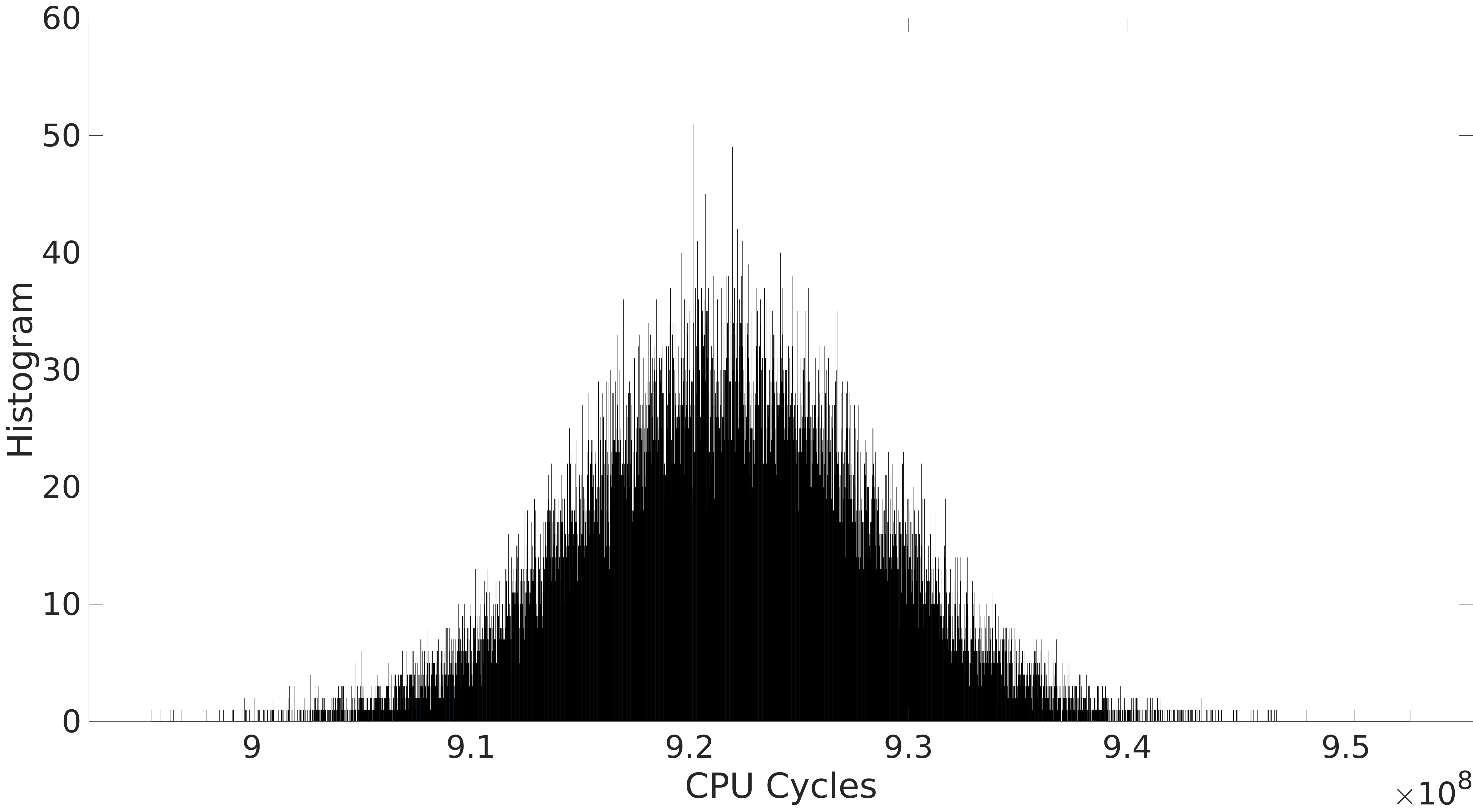}
 \caption{Histogram of RSA-2048 signature generation timings on a dedicated Infineon TPM as measured on a Core i7-8650U machine for 40,000 observations.}
     \label{fig:rsa_inf}
 \end{figure}

\begin{figure}[htp]
    \centering
\includegraphics[width=0.90\linewidth]{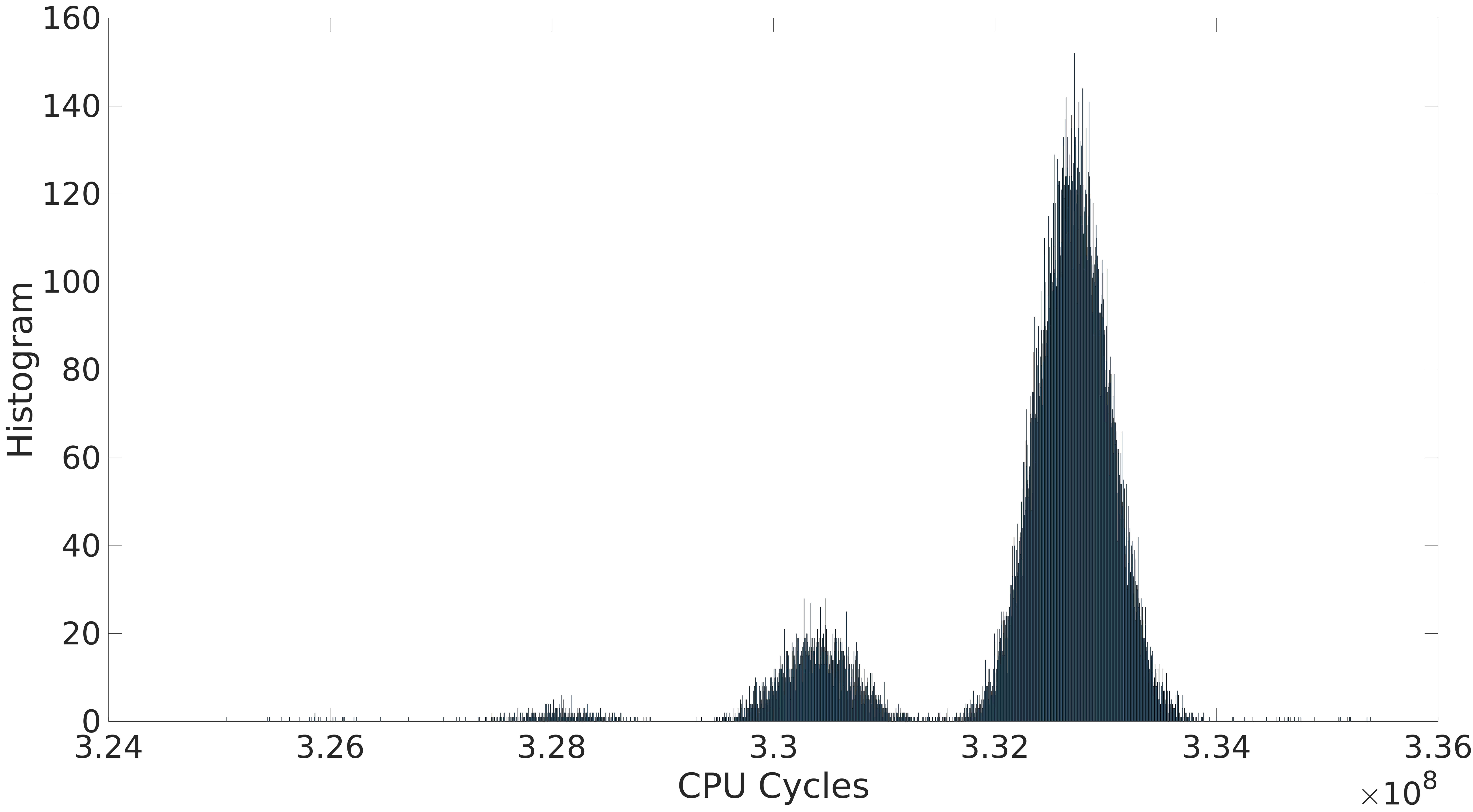} 
\caption{Histogram of ECSchnorr (NIST-256p) signature generation times on Intel \fTPM\ as measured on a Core i7-7700 machine for 34,000 observations.}
    \label{fig:histecschnorr}
\end{figure}

\begin{figure}[htp]
    \centering
\includegraphics[width=0.90\linewidth]{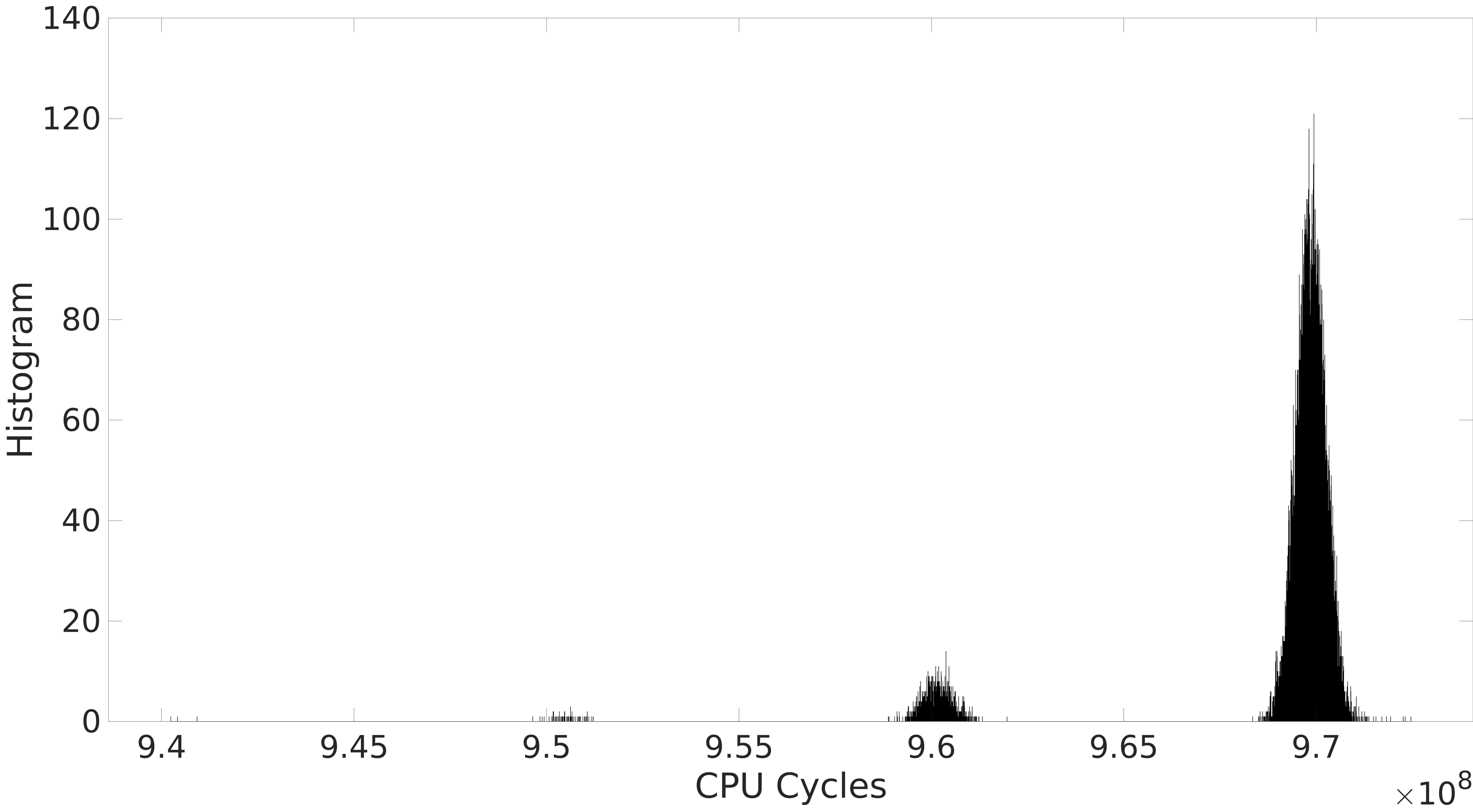}
\caption{Histogram of ECDSA (BN-256) signature generation times on Intel \fTPM\ as measured on a Core i7-7700 machine for 15,000 observations. Using the BN-256 curve approximately doubles the execution time of ECDSA, which makes the multiplication windows even more distinguishable.}
    \label{fig:histbn256}
\end{figure}

\begin{figure}[htp]
    \centering
\includegraphics[width=0.90\linewidth]{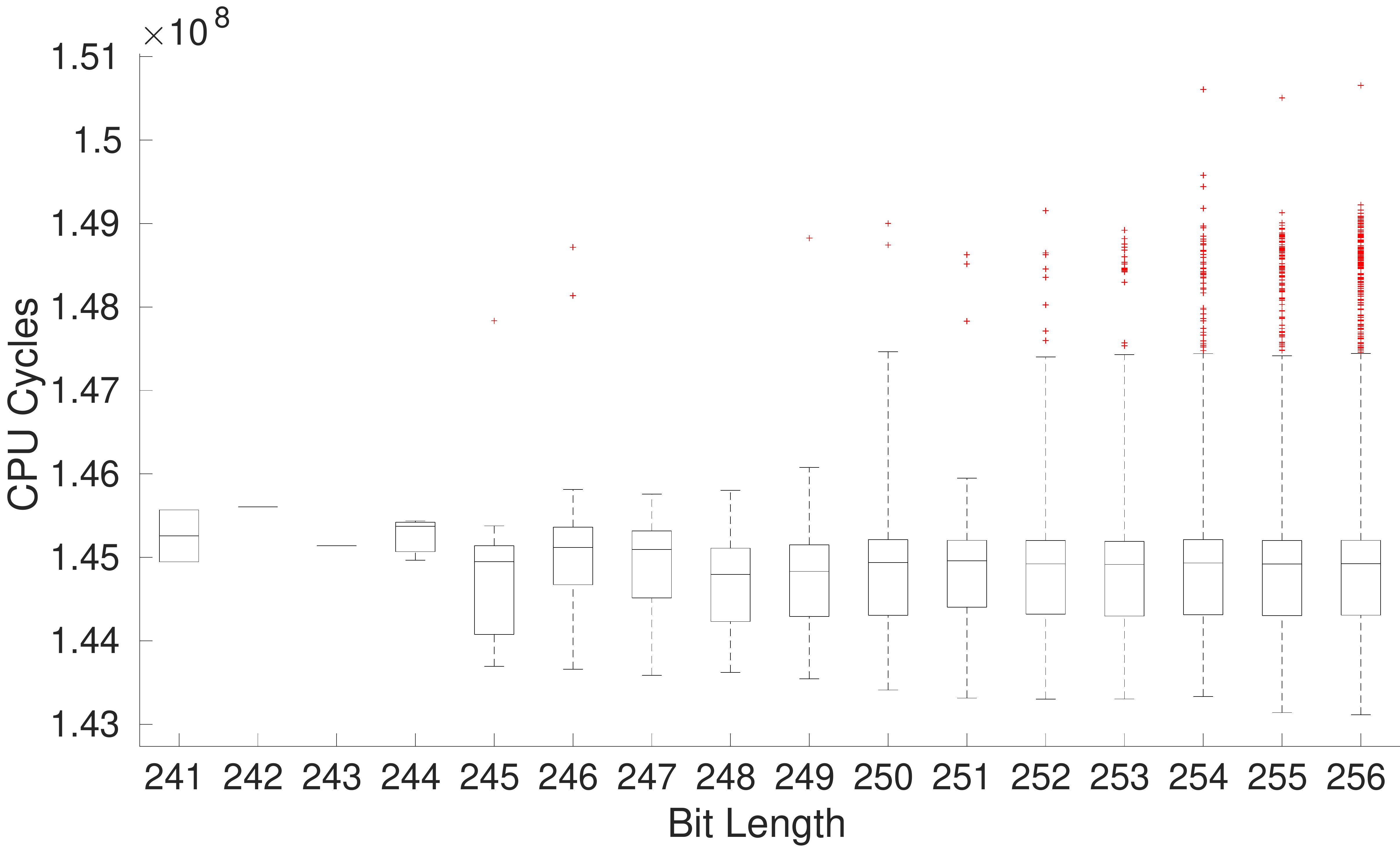} 
\caption{Histogram of ECDSA (NIST-256p) signature generation timings a dedicated Infineon TPM as measured on a Core i7-8650U machine for 40,000 observations.}
    \label{fig:ecdsa_inf_boxplot}
\end{figure}

\begin{figure}[htp]
    \centering
\includegraphics[width=0.90\linewidth]{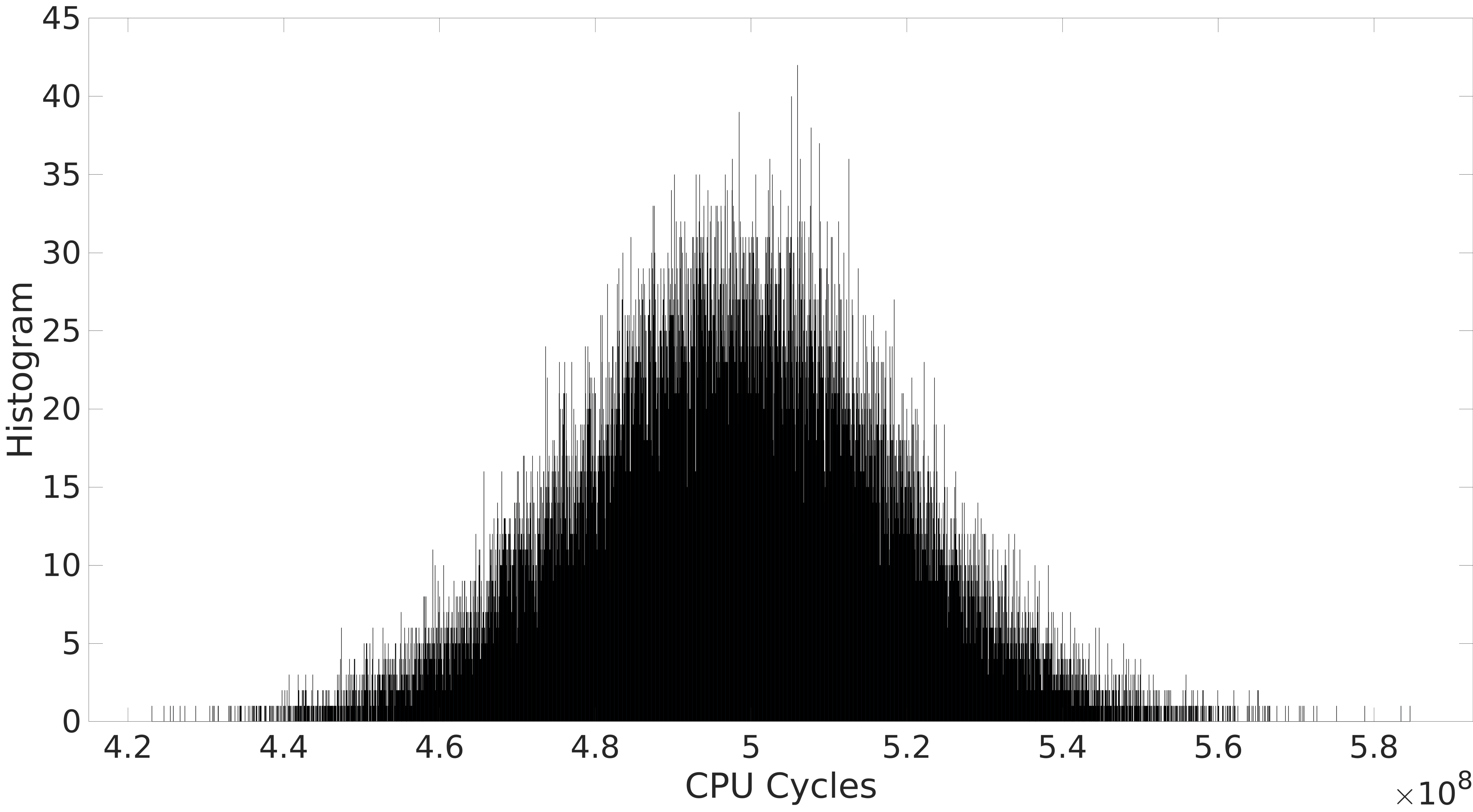} 
\caption{Histogram of ECDSA (NIST-256p) signature generation timings a dedicated Nuvoton TPM as measured on a Core i5-6440HQ machine for 40,000 observations.}
    \label{fig:ecdsa_nuv}
\end{figure}

\end{document}